\documentclass[useAMS,usenatbib]{mn2e}
\usepackage{myaasmacros}
\usepackage{graphicx}
\usepackage{amssymb}

\title[Orbits for the Milky Way's dwarfs]{Determining orbits for the Milky Way's dwarfs}

\author[H. Lux et al.]{H. Lux,$^1$\thanks{lux@physik.uzh.ch} J. I. Read$^{2}$, G. Lake$^1$ \\  $^1$ Department of Theoretical Physics, University of Z\"urich, Winterthurerstr. 190, CH-8057 Z\"urich, Switzerland\\ $^2$ Department of Physics \& Astronomy, University of Leicester, University Road, Leicester,  LE1 7RH, United Kingdom.}
 
\begin{document}

\date{Accepted ??. Received ??; in original form ??}

\maketitle

\begin{abstract}
We calculate orbits for the Milky Way dwarf galaxies with proper motions, and compare these to subhalo orbits in a high resolution cosmological simulation. We use the simulation data to assess how well orbits may be recovered in the face of measurement errors, a time varying triaxial gravitational potential, and satellite-satellite interactions. For present measurement uncertainties, we recover the apocentre $r_a$ and pericentre $r_p$ to $\sim 40$\%. With improved data from the Gaia satellite we should be able to recover $r_a$ and $r_p$ to $\sim 14$\%, respectively. However, recovering the  3D positions and orbital phase of satellites over several orbits is more challenging.  This owes primarily to the non-sphericity of the potential and satellite interactions during group infall.  Dynamical friction, satellite mass loss and the mass evolution of the main halo play a more minor role in the uncertainties.

We apply our technique to nine Milky Way dwarfs with observed proper motions. We show that their mean apocentre is lower than the mean of the most massive subhalos in our cosmological simulation, but consistent with the most massive subhalos that form before $z=10$. This lends further support to the idea that the Milky Way's dwarfs formed before reionisation. 
\end{abstract}

\begin{keywords}
galaxies: dwarf, galaxies: kinematics and dynamics, methods: numerical 
\end{keywords}

%__________________________________________________________

\section{Introduction}\label{sec:intro}
The dwarf galaxies of the Milky Way provide a unique window onto galaxy formation on the smallest scales. Their close proximity allows us to resolve individual stars, obtaining detailed star formation histories, mass measurements, proper motions, and distances \citep{1998ARA&A..36..435M}. Increasingly accurate proper motions present the possibility of reliable orbit calculations for the dwarfs \citep[e.g.][]{1995AJ....110.2747S,1996AAS...188.0901S,2002AJ....124.3198P,2003AJ....126.2346P,Dinescu:2004pe,2005AJ....130...95P,2006ApJ...652.1213K,2009AJ....137.4339C}. We may then explore links between environment and formation history, e.g. the morphology-distance relation [\citep{1994AJ....107.1328V} although recent discovery of three dSph at large distances from Andromeda and the Milky Way calls this into question \citep[e.g.][]{2008ApJ...688.1009M}]. With accurate 3D determinations of their orbits, we may then also determine if the dwarfs fell into the Milky Way in groups \citep{2008ApJ...686L..61D,2008MNRAS.385.1365L,2009arXiv0909.1916K} or suffered past interactions \citep{2007MNRAS.379.1475S,2008ApJ...675..201M}. 

In this paper, we derive orbits for the Milky Way's dwarfs with proper motion data, by integrating their motion backwards in a static gravitational potential. We estimate the effects of several systematic errors by using a similar integration of subhalos in the high resolution Via Lactea I cosmological simulation of a Milky Way analogue \citep[VL1; ][]{2007ApJ...657..262D}. We assess the impact of measurement errors by adding realistic errors to the subhalos' redshift $z=0$ phase space position, and tracing their orbits back in a fixed potential.  Our fiducial model integrates orbits in a fixed spherical potential; however we also examine the effects of a triaxial potential, mass evolution of the main halo, dynamical friction, and mass loss from the satellites. We devise three measures of quality for tracing back orbits, of increasing difficultly. These are the comparison of true and recovered: 

\begin{enumerate}
\item last pericentre $r_p$ and apocentre $r_a$
\item $r_p$, $r_a$ and the orbital period $t$ backwards in time over $N$ orbits. 
\item  {\it 3D} pericentre ${\bf r_p}$ and apocentre ${\bf r_a}$ backwards in time over $N$ orbits.
\end{enumerate} 

The first of these three tells us the current orbit of the dwarf for comparison with cosmological simulations. It represents the minimum useful information we might obtain from a dwarf orbit. The second tells us {\it when} the dwarf encountered the Milky Way, and how close it came. This allows us to compare orbits with star formation histories. The third tells us the full 3D position of the dwarf as a function of time. This tells us if dwarfs ever interacted via group infall or fast flyby.

This paper is organised as follows. In \S\ref{sec:tests}, we examine whether orbits can be recovered  at levels (i), (ii) and (iii) within the model limitations and what data quality is required for this. In \S\ref{sec:data}, we apply our method to nine Milky Way dwarf galaxies with measured radial velocities and proper motions. Finally, in \S\ref{sec:conclusions} we present our conclusions. 

%__________________________________________________________
\section{Testing the method}\label{sec:tests}

In this section, we use the Via Lactea I (VL1) simulation of a Milky Way mass galaxy \citep{2007ApJ...667..859D}, to determine how well we recover satellite orbits in the face of measurement errors and a time varying gravitational potential. We extract from the simulation three sets of subhalos\footnote{We refer to subhalos as `satellites' interchangeably.}: the 50 most massive today ($z^{50}_0$), the 50 most massive before redshift $z=10.59$ ($z^{50}_{10}$) and the 50 most massive before redshift $z=10.59$, taking depletion by a disc into account ($z^{50}_{10}(r_{d}))$\footnote{See the data at http://www.ucolick.org/~diemand/vl/ }. In all cases, we include only subhalos with mass $M > 10^7 M_\odot$ and distance to the centre of the main halo $r< 150$\,kpc at redshift $z=0$. This represents the mass and radius range where we find the Milky Way dwarfs (c.f. Table \ref{tab:dwarfdata1} in section \S\ref{sec:thedata}). In the disc depleted sample we exclude all orbits having pericentres $r_p<r_d$ before extracting the sample, motivated by recent work by \citet{2009arXiv0907.3482D}. We use $50$ subhalos to create a sensible sized sample that likely corresponds to the expected number that might be observed with full sky coverage at a depth comparable to the Sloan Digital Sky Survey \citep{2007ApJ...670..313S,2008ApJ...686..279K,2009AJ....137..450W}.

We take the present day phase space position of these satellites and integrate them backwards in time to compare our derived orbits with the true VL1 orbits. We use a Leapfrog time integrator with adaptive time stepping to adjust for the higher resolution required at pericentre \citep[e.g.][]{1992nrca.book.....P}. We scale each time step by a fraction of the instantaneous orbital time, similarly to \citet{2007MNRAS.376..273Z}. The apocentre $r_a$ and pericentre, $r_p$ of each orbit (and therefore the period $t$) are then recovered by searching for a sign change in $dr/dt$, where $r$ is the distance from the satellite to its host galaxy. For the VL1 data, we use a more sophisticated algorithm, that only characterises global extrema in the orbit around the main halo and ignores local extrema caused by satellite-satellite interactions or the satellite behaviour before falling into the main halo. We explicitly exclude the cosmological turn around as part of the real orbit. Note that for both the simulation as well as the recovered orbits, we determine half periods $t_{1/2}$, i.e. the temporal distance between two orbital extrema. Hence, we only include orbits with 2 orbital extrema.  By this criteria any satellites with long periods are excluded, which are falling in for the first time. Whenever we write period instead of half period, we mean the half period multiplied by two.

\begin{figure}
\centering
\includegraphics[width=0.23\textwidth]{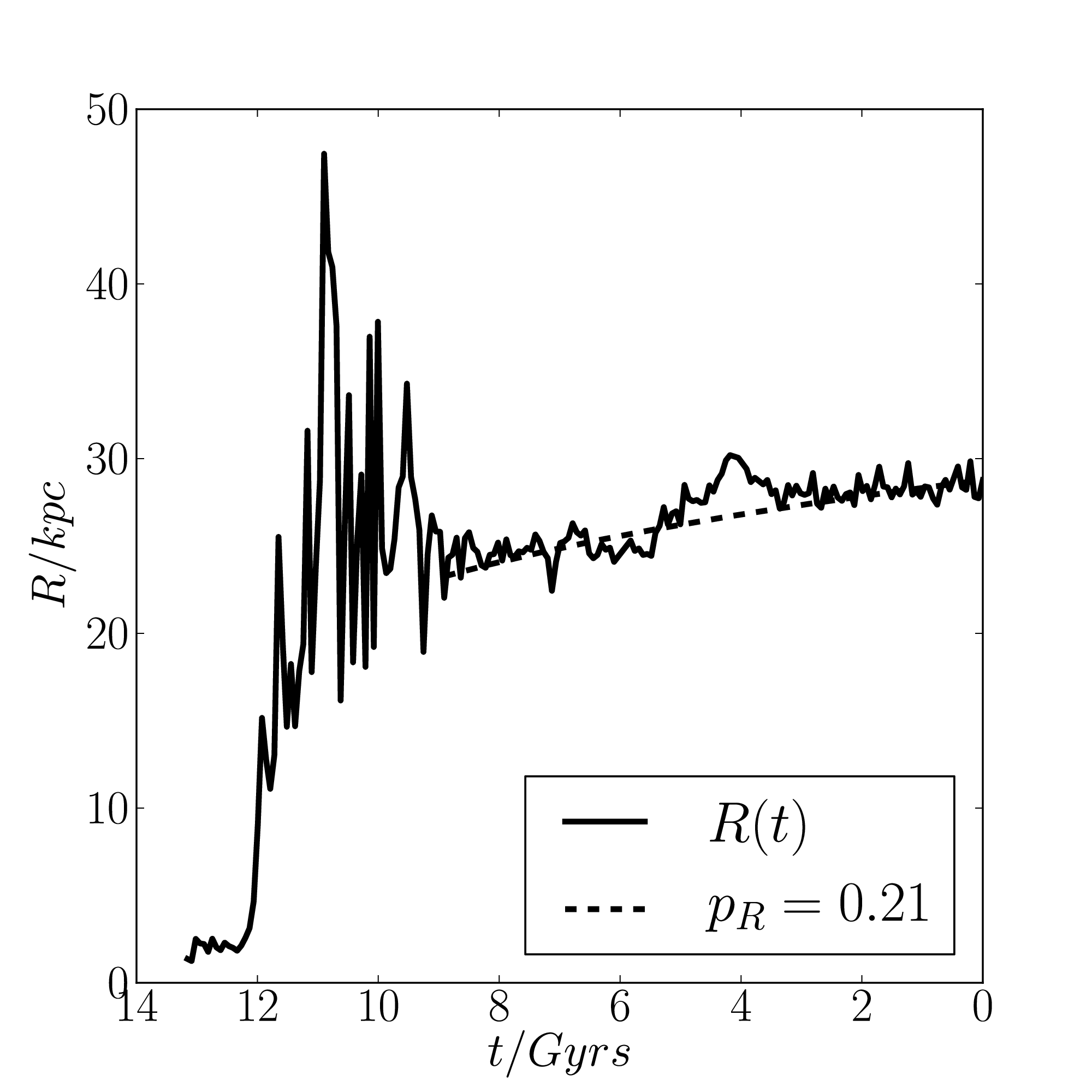}
\includegraphics[width=0.23\textwidth]{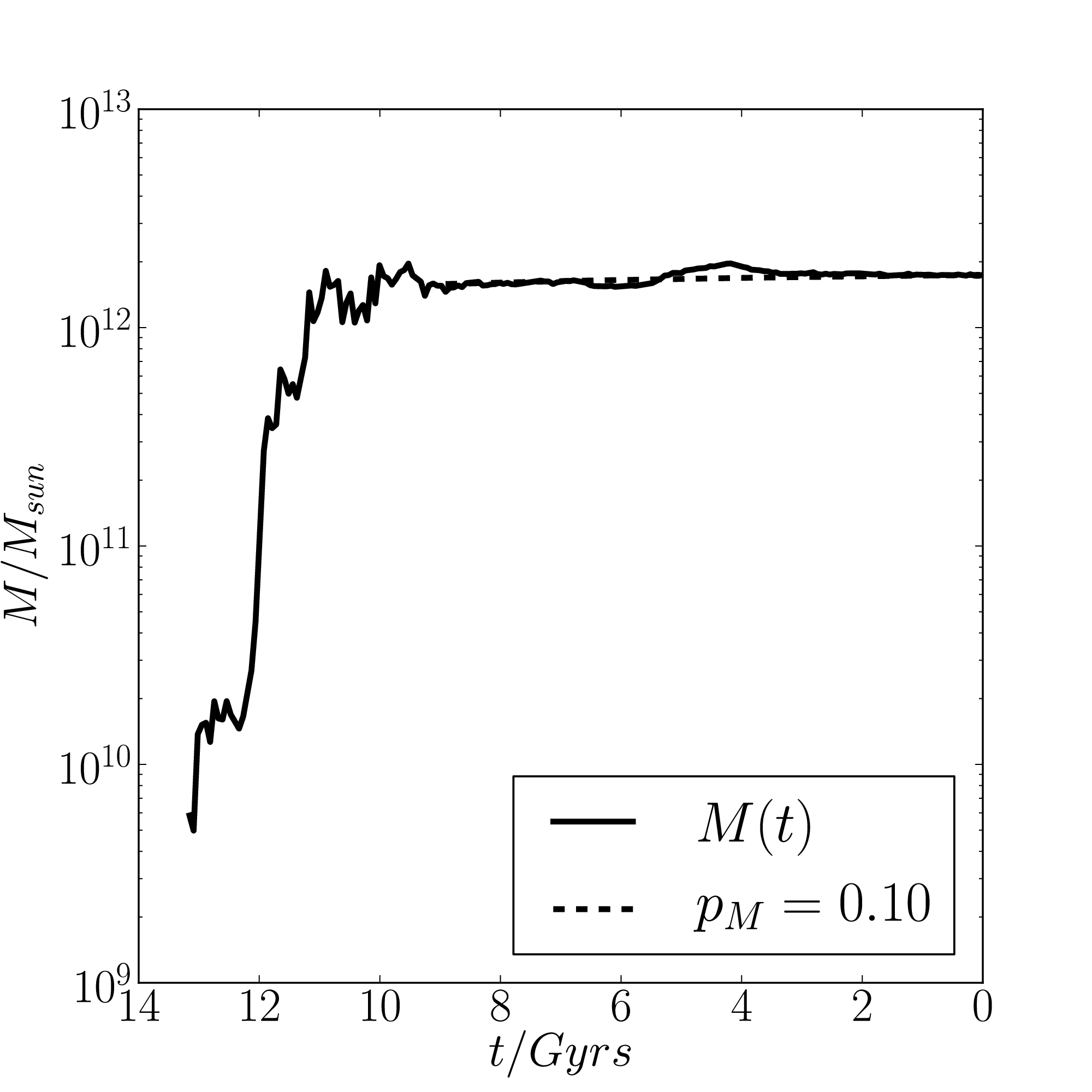}
\caption{Evolution of the mass and scale radius of the VL1 main halo (solid line). The scale radius is more difficult to constrain for high redshifts. The dashed line is the best fitting power law going back 9\,Gyrs in time.}
\label{fig:VLevol}
\end{figure}

\subsection{The orbit recovery models}\label{sec:orbmodels} 

We consider the following models for our orbit integration:

\begin{itemize}
\item {\bf The fiducial model ($F$)} uses a static, spherical, NFW potential \citep{1996ApJ...462..563N}. In $F$, we take parameters from the best fit to the VL1 main halo at redshift $z=0$: $M_{200}=1.77\times10^{12}\,$M$_\odot$, $R_{200}=389$\,kpc and $r_c=24.6$\,kpc \citep{2007ApJ...657..262D}.

\item {\bf The dynamical friction model ($DF$)} uses the same potential as in model $F$ with dynamical friction forces added \citep{1943ApJ....97..255C}. We use a radially varying Coulomb logarithm as in \citet{2003ApJ...582..196H}. We also explore the impact of mass loss of the infalling satellites, and the mass growth of the VL1 halo with time (see Figure \ref{fig:VLevol}). We fit simple functions to the subhalo mass loss history, and VL1 main halo mass growth with time as in \cite{2004MNRAS.351..891Z}. 

\item {\bf The triaxial model ($T$)}: This uses a triaxial NFW potential as in \citet{2007ApJ...671.1135K} and \citet{2009ApJ...702..890G}. We use axis ratios: $q=0.83$ and $s=0.8$, as measured for the VL1 main halo at outer radii at $z=0$.

\item {\bf The double halo mass model ($2M$)}: This uses the potential as in model $F$, but doubling the mass of the main halo (doubling the scale radius has a negligible effect).

\end{itemize}

These different models allow us to assess the importance of systematic effects on our orbit recovery for real Milky Way dwarf data. In practice, only model $F$ can be realistically applied to real data since we do not know the mass loss history of the Milky Way dwarfs, nor the mass growth history or shape of the Milky Way halo. We use models $DF$, $T$ and $2M$ to explore the systematic impact of this poor knowledge on our orbit recovery. 

%__________________________________________________________

\subsection{Results}\label{sec:testresults}

In this section, we measure how well we can recover the orbits of subhalos in our VL1 data set. We use each of the the recovery models presented in \S\ref{sec:orbmodels}, with and without measurement errors and assess our results using criteria (i), (ii) and (iii) in \S\ref{sec:intro}. 

\begin{figure*}
\centering
\includegraphics[width=0.4\textwidth]{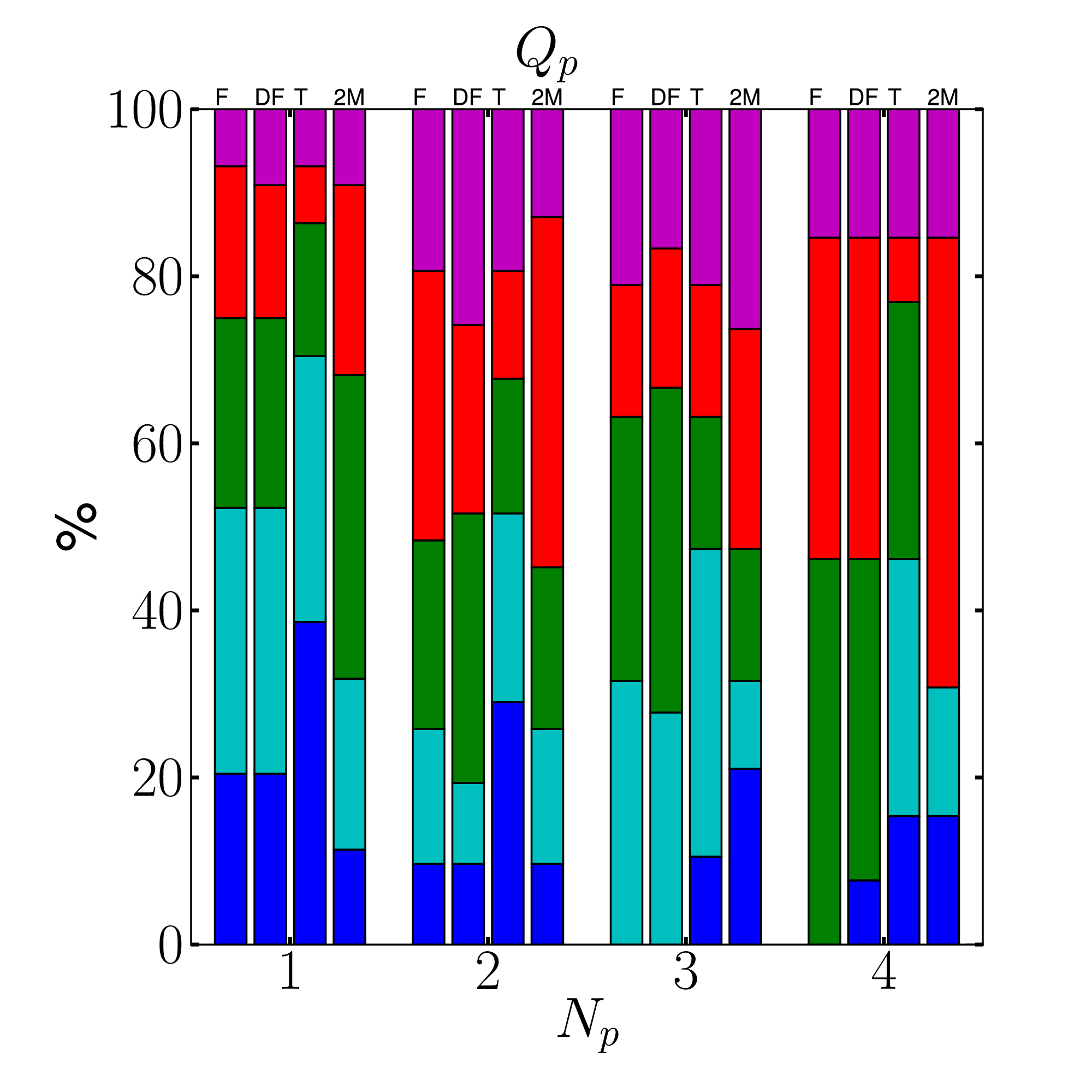}\hspace{-3mm}
\includegraphics[width=0.4\textwidth]{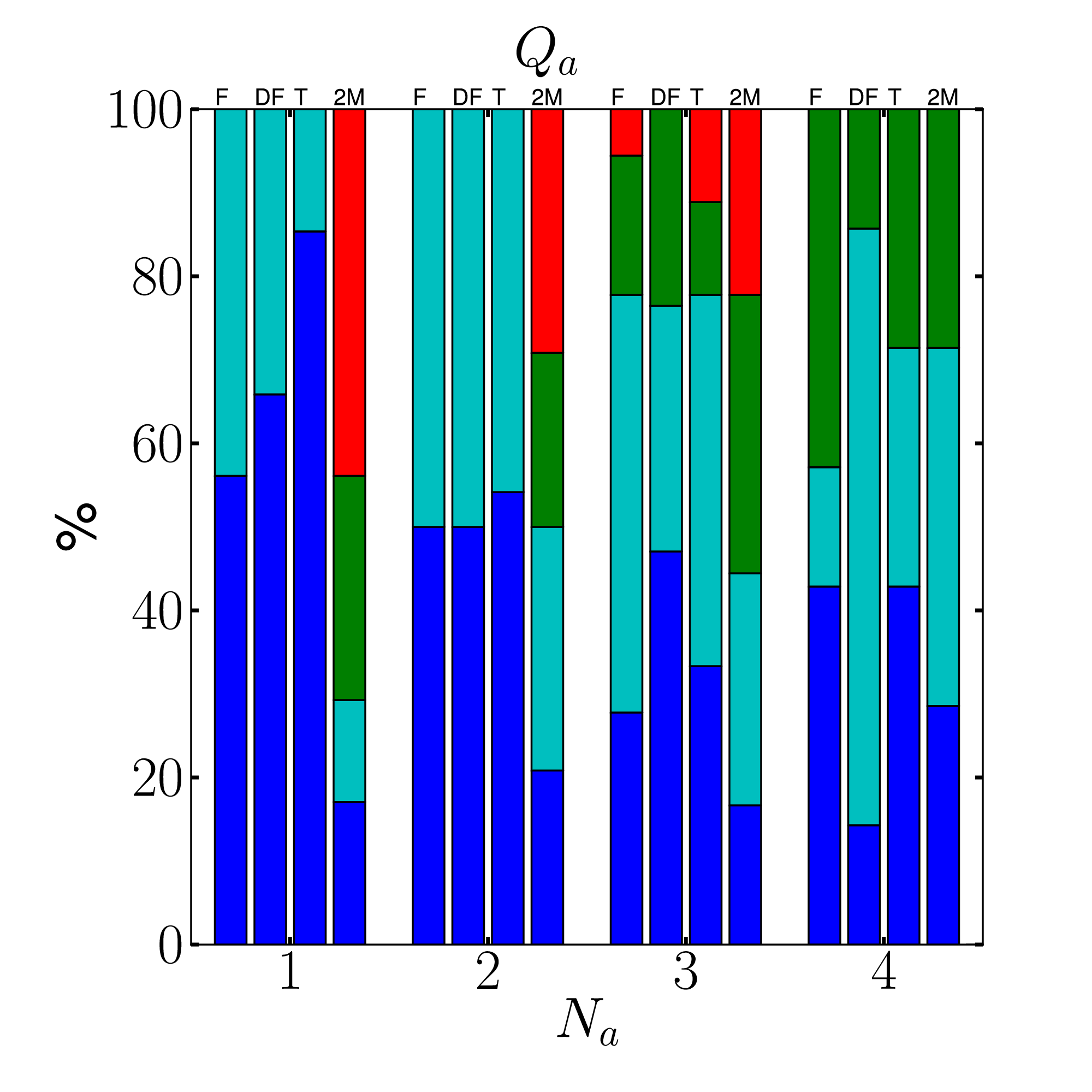}\hspace{-3mm}
\caption{Orbit recovery in models $F$, $DF$, $T$ and $2M$  with no measurement errors. The coloured bars show the percentage of subhalos with fractional error $Q = 0-0.1$ (blue), $0.1-0.3$ (cyan), $0.3-0.5$ (green), $0.5-0.8$ (red) and $>0.8$ (magenta) over $N$ orbits backwards in time. In our Fiducial model ($F$), 20\% of satellites have their most recent pericentre recovered to better than 10\%, while 56\% have their most recent apocentre recovered to better than 10\%.}
\label{fig:nomeasure1d}
\end{figure*}

\begin{figure*}
\centering
\includegraphics[width=0.4\textwidth]{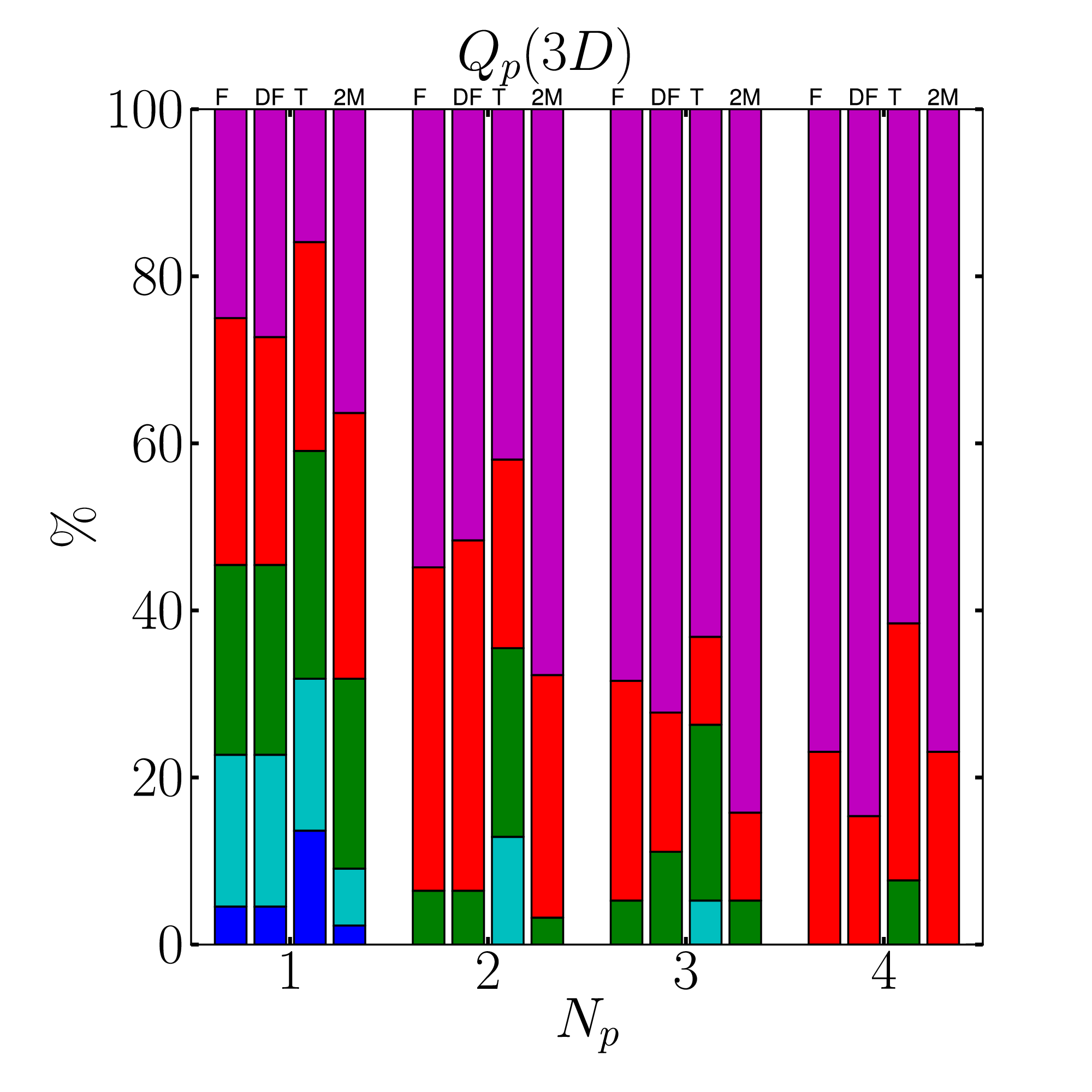}\hspace{-3mm}
\includegraphics[width=0.4\textwidth]{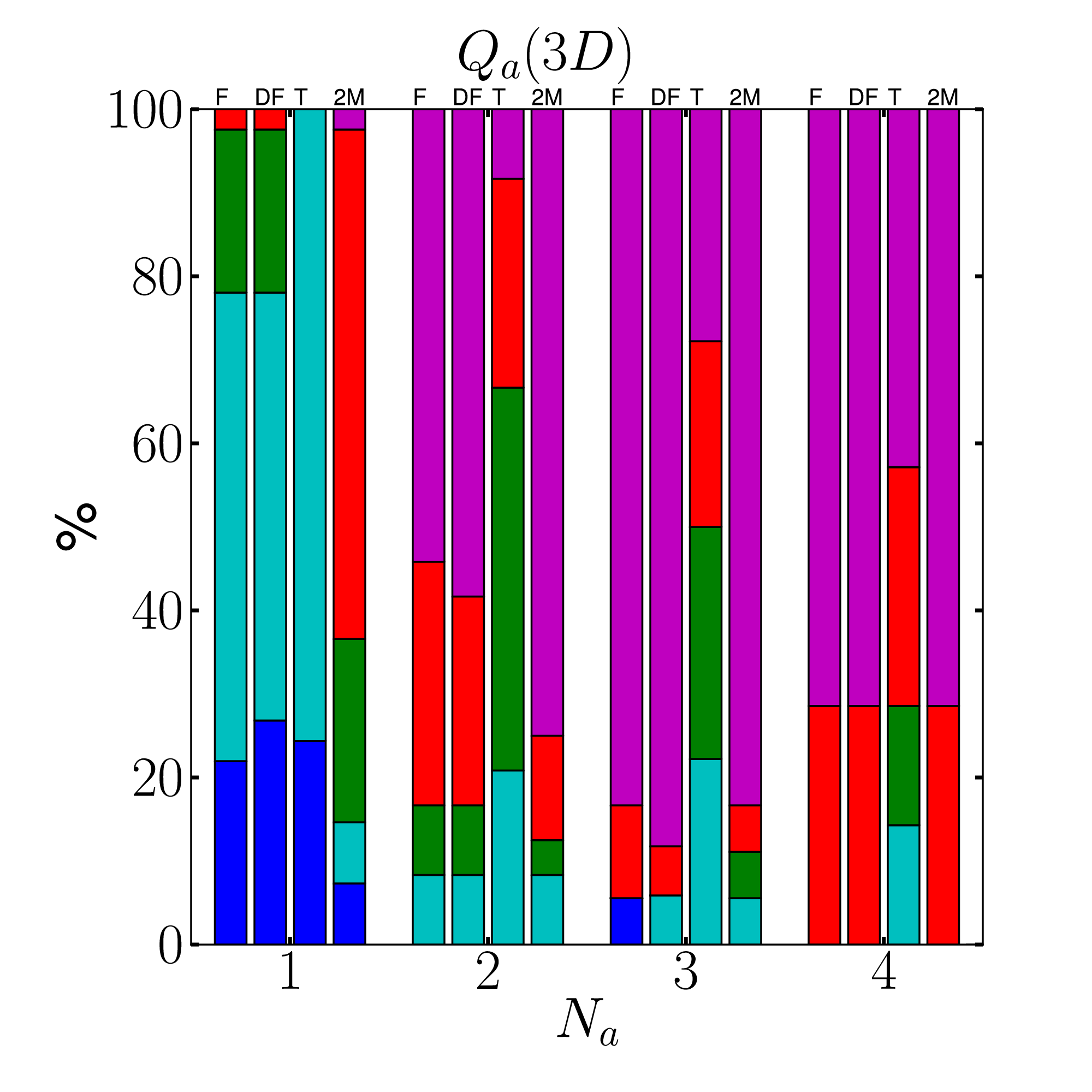}\hspace{-3mm}
\caption{As Figure \ref{fig:nomeasure1d}, but for the 3D pericentre (left) and 3D apocentre (right).} 
\label{fig:nomeasure3d}
\end{figure*}

\subsubsection{Perfect data}\label{sec:perfect} 

To assess the model systematics, we first consider the case without position and velocity measurement uncertainties. 
Figures \ref{fig:nomeasure1d}, \ref{fig:nomeasure3d} and \ref{fig:nomeasureT} show our recovery for each of the models $F, DF, T$ and $2M$ for the `1D' pericentres $r_p$ and apocentres $r_a$, the 3D pericentres ${\bf r_p}$ and apocentres ${\bf r_a}$ and the half period $t_{1/2}$ over $N$ orbits. The coloured bars show the percentage of subhalos with fractional error $Q = 0-0.1$ (blue), $0.1-0.3$ (cyan), $0.3-0.5$ (green), $0.5-0.8$ (red) and $>0.8$ (magenta). We define the relative error $Q$ as:

\begin{equation}
Q_p = \frac{r_p - r_{p,t}}{r_{p,t}} \hspace{2mm} ; \qquad Q_p(3D) = \frac{|{\bf r_p} - {\bf r_{p,t}}|}{|{\bf r_{p,t}}|}
\end{equation} 
for the `1D' pericentre (left) and 3D pericentre (right), where $r_{p,t}$ is the true pericentre and $r_p$ is the recovered pericentre. $Q$ is defined similarly for the other orbit quantities $r_a$, ${\bf r_a}$ and $t_{1/2}$. Note that in comparing $r_p$ and $r_{p,t}$, the limited time resolution of the VL1 simulation can lead to an additional error (that causes us to always over estimate $r_{p,t}$). We tested the effect of this, finding that only orbits with $r_p \lesssim 20$\,kpc are affected. The additional error due to finite time resolution is small compared to our model uncertainties.

\subsubsection{Recovering the 1D pericentre and apocentre} 

From Figure \ref{fig:nomeasure1d}, we see that the last apocentre is well recovered in our fiducial model $F$. No satellites have their last apocentre recovered with greater than 30\% error. This is perhaps to be expected. Subhalos are most likely to be at apocentre now and so their present position already gives a reasonable constraint on $r_a$. The $DF$ and $T$ models both improve the recovery for the most recent apocentre. This suggests that the primary reason for our error in the most recent apocentre recovery is a combination of wrong halo shape and dynamical friction, where the halo shape is more significant. These effects will be swamped by other effects for the apocentre recovery further back in time (c.f. \ref{sec:group}). By contrast, model $2M$ gives a poor performance, with only $\sim 20$\% of the subhalos having better than 10\% error on the last apocentre. This demonstrates that it is critical to have a good estimate of the mass of the Milky Way interior to the orbits of its dwarfs. 

The last pericentres are significantly less well recovered than the apocentres, with only 20\% having an error better than 10\% in our model $F$. Here the halo shape plays a major role, with model $T$ giving a significantly better performance ($\sim40$\% recovered to better than 10\%); dynamical friction (model $DF$) has little impact on the recovery. To better understand why the pericentre recovery fails, some typical example orbits from each $Q_p$ bin are given in Figure \ref{fig:QcompP}. As can be seen, orbits that have a well recovered last $r_p$ are {\it short period} orbits. These are well recovered in model $F$ and well recovered in the triaxial model $T$ that has the correct current halo shape. However, long period orbits sample the potential over several  giga years backwards in time. Here even our triaxial model (fit to the VL1 halo at redshift $z=0$) fails and the $r_p$ recovery is poor (see the magenta orbit). 

\subsubsection{recovering orbits over several periods: The problem of group infall}\label{sec:group}

Recovering orbits over several periods backwards in time is more challenging. The fraction of orbits with better than 10\% error in $r_a$ in our triaxial model $T$ drops by $\sim40$\% by $N_a=2$; while there are no well recovered $r_p$ in our model $F$ by $N_p=3$. The triaxial model $T$ gives an improvement in the recovered $r_p$, but no improvement in $r_a$. Note that there are some fluctuations in this trend due to small number statistics in the $N_a=3$ and $N_a=4$ bins. So individual orbits, that are improved in one of the models have a higher impact than they would have in a larger sample.

\begin{figure}
\centering
\includegraphics[width=0.4\textwidth]{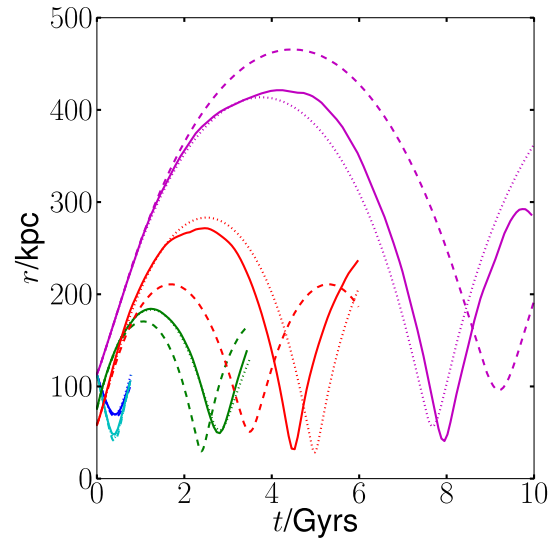}
\caption{True and recovered orbits for the first pericentre with different $Q_p$ values. The colours correspond to the bins in Figure \ref{fig:nomeasure1d}: $Q_p = 0-0.1$ (blue), $0.1-0.3$ (cyan), $0.3-0.5$ (green), $0.5-0.8$ (red) and $>0.8$ (magenta). Solid lines are the true orbit; dashed lines are the recovered orbits in our fiducial model $F$; and dotted lines are the recovered orbits in our triaxial model $T$. } 
\label{fig:QcompP}
\end{figure}

\begin{figure}
\centering
\vspace{3mm}
\includegraphics[width=0.4\textwidth]{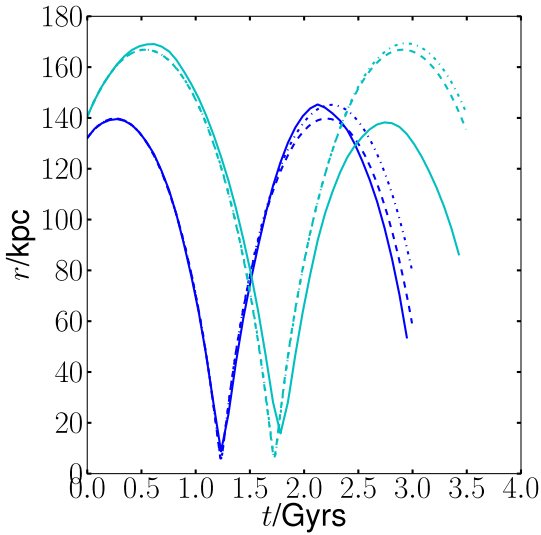}
\caption{True and recovered orbits for the second apocentre ($N_a=2$) with different $Q_a$ values. The colours correspond to the bins in Figure \ref{fig:nomeasure1d}: $Q_a = 0-0.1$ (blue), $0.1-0.3$ (cyan). Solid lines are the true orbit; dashed lines are the recovered orbits in our fiducial model $F$; and dash-dotted lines are the recovered orbits in our dynamical friction model $DF$.} 
\label{fig:QcompA}
\end{figure}

To understand why the errors on $r_a$ grow so rapidly backwards in time, we give some example orbits recovered at $N_a=2$ in Figure \ref{fig:QcompA}. As with the $r_p$ (Figure \ref{fig:QcompP}), the longer period orbits are more poorly recovered. 

\begin{figure*}
\centering
\includegraphics[width=0.3\textwidth]{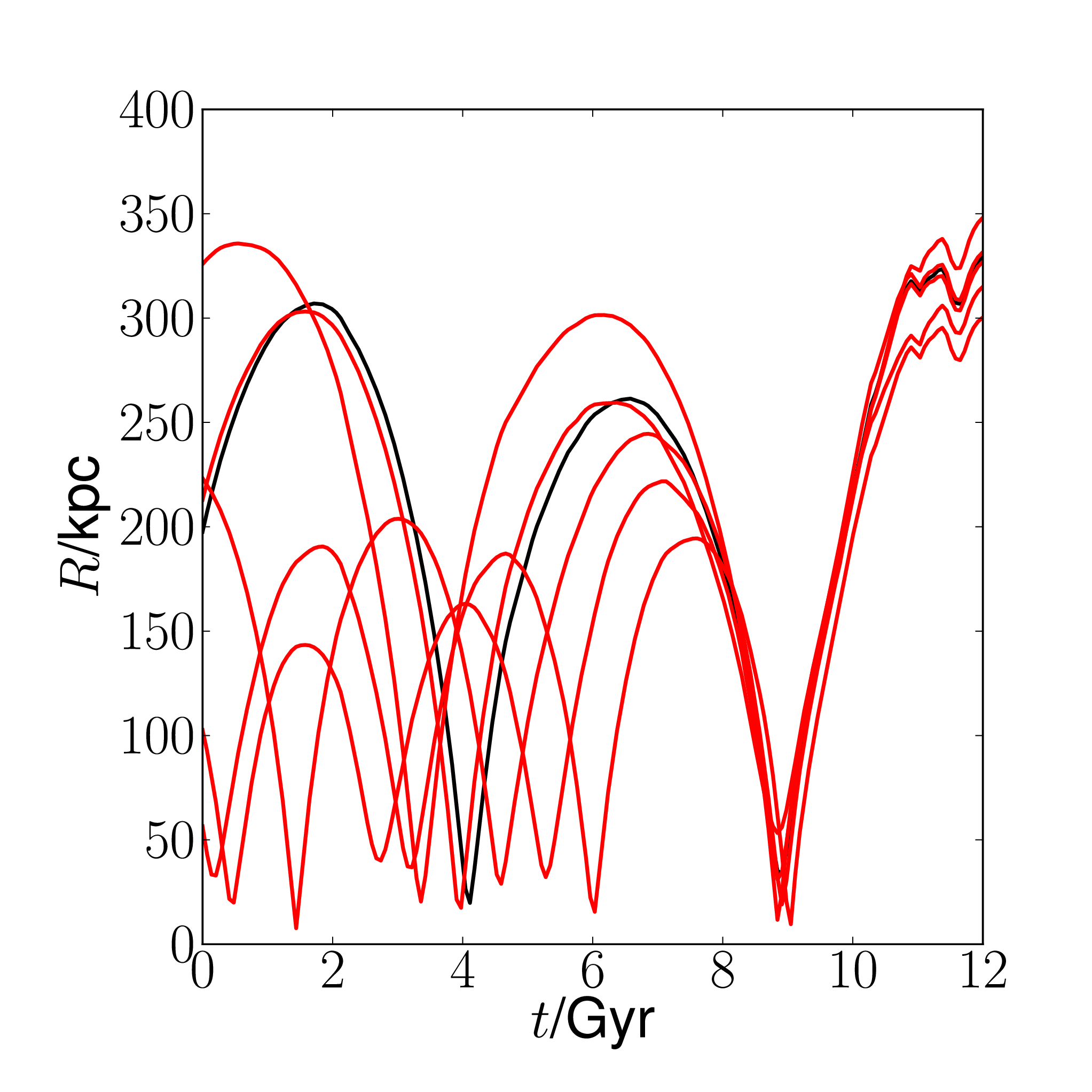}
\includegraphics[width=0.3\textwidth]{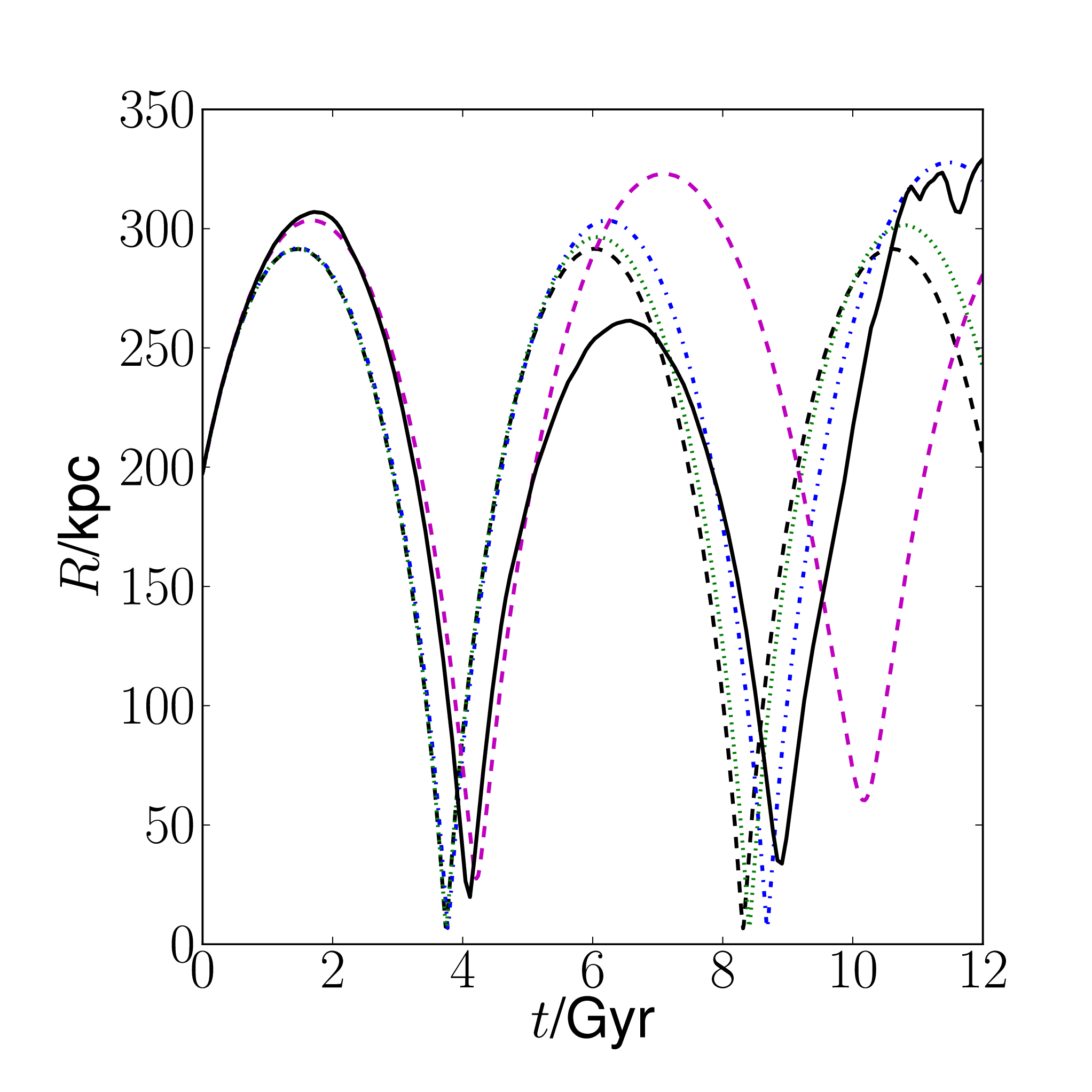}
\includegraphics[width=0.3\textwidth]{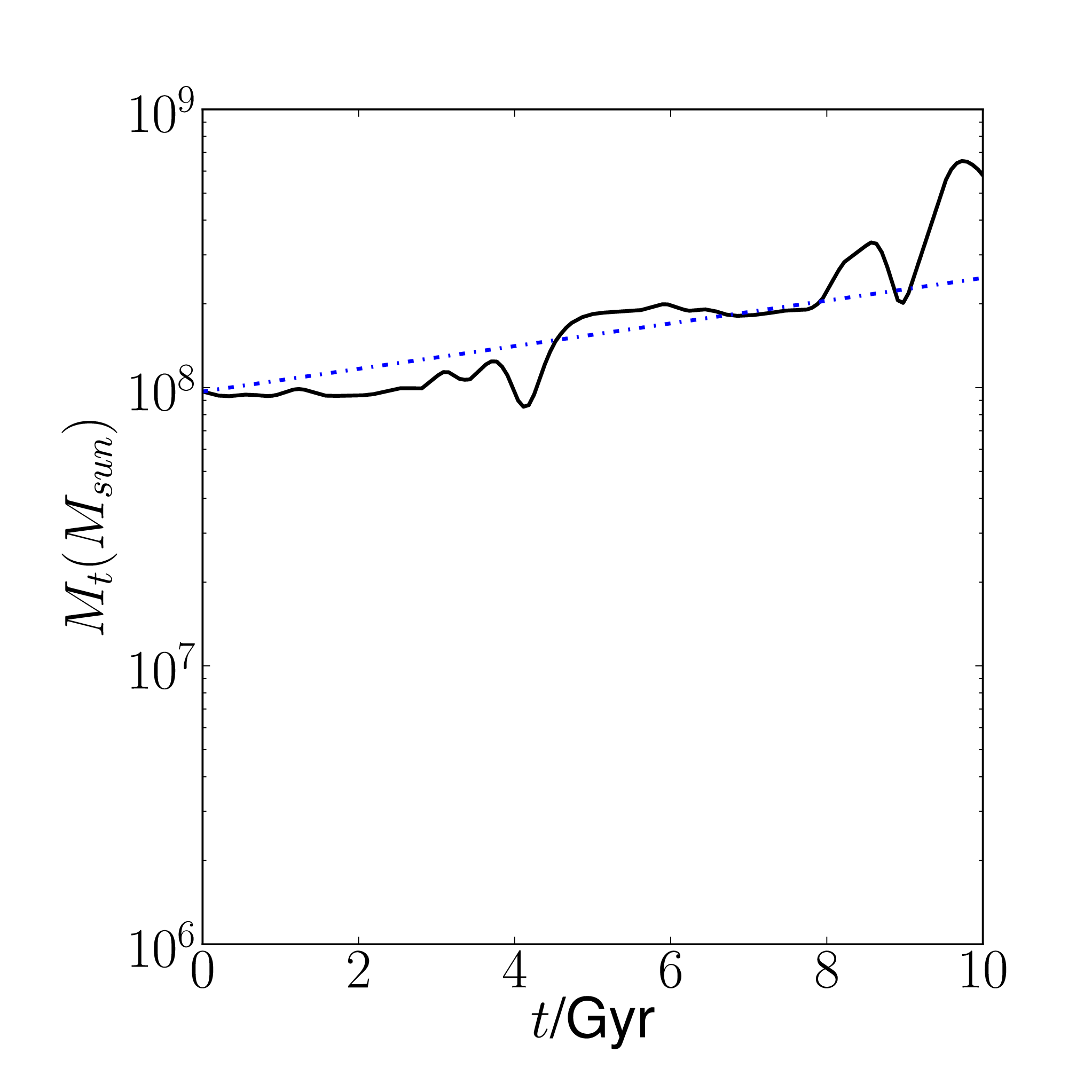}
\caption{Group infall in the VL1 simulation. The left panel shows the orbits of subhalos that fall in inside a group (red and black lines); the most massive member is shown in black. The middle panel shows our orbit recovery for the most massive subhalo in this group in model $F$ (black dashed line), $T$ (magenta dashed line), $DF$ (blue dash-dotted line) and an orbit integrated with dynamical friction, but without mass loss of the satellite nor mass evolution of the main halo (green dotted line). The right panel shows the mass evolution of the most massive subhalo in the group and our best fit curve (blue dot-dashed line) used in model $DF$.} 
\label{fig:Group}
\end{figure*}

There is a significant class of orbits -- 50\% -- with $0.1 < Q_a<0.3$ (cyan) that are less well recovered at $N_a=2$.  We explore a typical orbit in this class in more detail in Figure \ref{fig:Group}. The middle panel shows our orbit recovery for the most massive subhalo in this group in model $F$ (black dashed line), $T$ (magenta dashed line), $DF$ (blue dash-dotted line) and an orbit integrated with dynamical friction, but without mass loss of the satellite nor mass evolution of the main halo (green dotted line). Notice that none of our models give a good fit to the orbit. Accounting for dynamical friction and mass loss (see right panel) as well as using the triaxial potential significantly {\it overestimates} the apocentre. We investigated this orbit further and find this satellite to be part of a loose group. The left panel shows the orbit of the satellite (black) and all subhalos that were initially close in phase space to this satellite (red). This grouping was found by determining all satellites closer than $4r_t$, where $r_t$ is the tidal radius defined in \citet{2007ApJ...657..262D}, for at least six time outputs. These `loose groups' usually break up at first pericentre, which causes an energy change that results in lower 2nd apocentre than we would expect from any of our integration models.

75\% of the the satellites in the cyan bin were found to be falling into the galaxy as part of a group. This suggests that the group infall statistics recently determined by \citet{2008MNRAS.385.1365L} may be a lower bound. We defer a detailed analysis of the statistics of loose group infall to future work, but note here that it appears to be responsible for many of our poor orbit determinations at $N_a\geq2$.

\subsubsection{recovering the orbital phase} 

\begin{figure}
\centering
\includegraphics[width=0.4\textwidth]{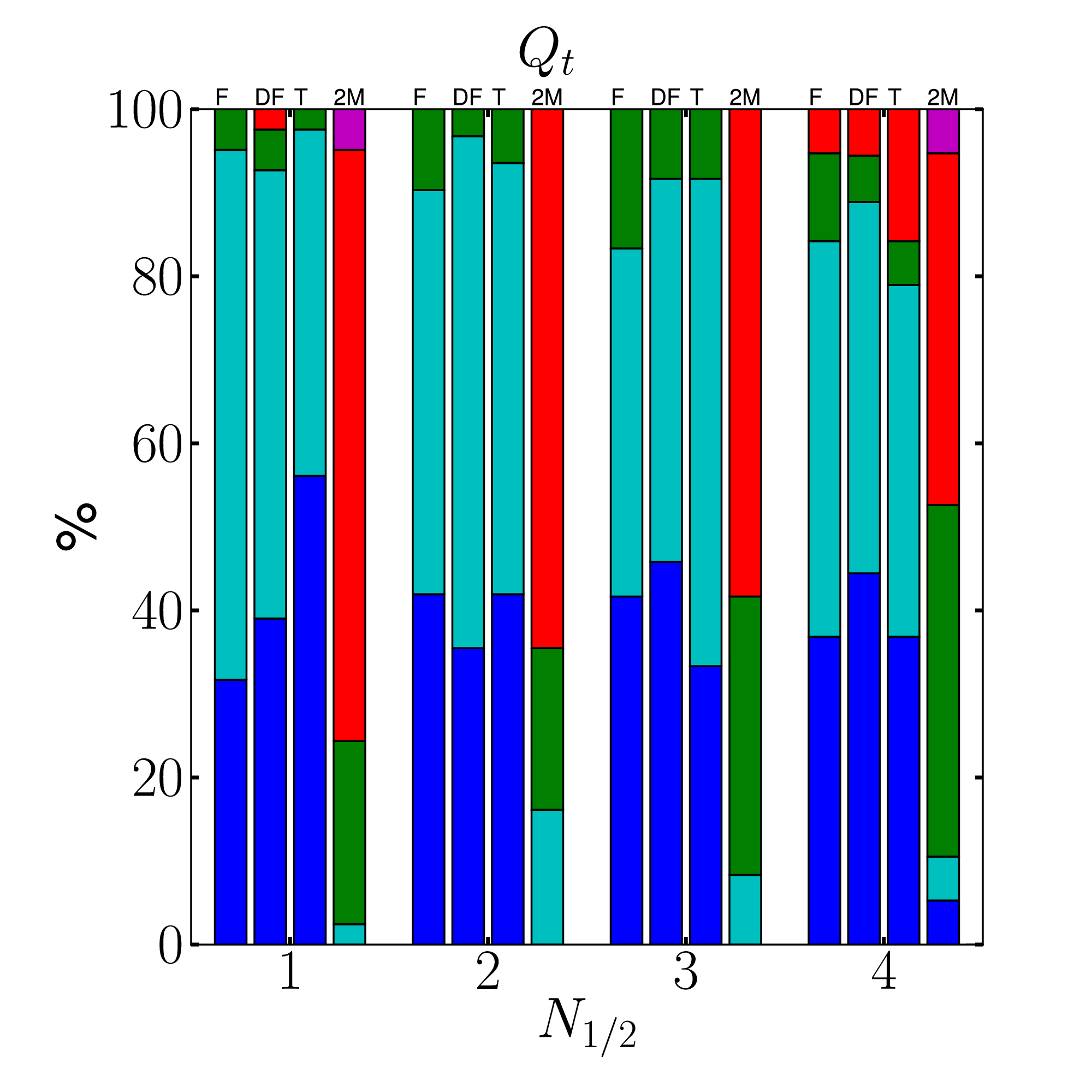}
\caption{As Figure \ref{fig:nomeasure1d}, but for half the orbital period $t_{1/2}$.} 
\label{fig:nomeasureT}
\end{figure}

In Figure \ref{fig:nomeasureT}, we show how well we recover the half orbital period over $N_{1/2}$ orbits assuming perfect data. Notice that the period is well-recovered up to even two orbits backwards in time. However, in determining the {\it phase} of the orbit, such period errors accumulate. This makes it challenging to try and match pericentre passages with observed star formation histories for the dwarfs. By $N_{1/2}=2$, our typical phase error is 0.6\,Gyrs, at $N_{1/2}=3$ it is already 0.8\,Gyr. Recall that this is for {\it perfect data} that have no measurement errors. With current proper motion errors, recovering the orbital phase is simply not possible (see \S\ref{sec:measure}).

\subsubsection{recovering the 3D pericentre and apocentre} \label{sec:3D}

Figure \ref{fig:nomeasure3d} shows how well we recover the 3D pericentre and apocentre for perfect data. Here having an accurate halo shape is vital. Our results in model $F$, even for a single orbit, are poor. Even in our triaxial model $T$ the results, while dramatically improved, are not encouraging. This is most likely due to the radial and temporal variations in the triaxiality as found by \citep{2007ApJ...671.1135K,2009ApJ...702..890G}. $\sim$20\% of the most recent 3D apocentre distances ${\bf r_a}$ and less than 5\% of the most recent 3D pericentre distances ${\bf r_p}$ are recovered to better than 10\% accuracy. Our results suggest that, unless the Milky Way halo is close to spherical or axisymmetric, recovering full 3D orbits for the Milky Way dwarfs will not be possible even with perfect data. This will make it difficult to determine if any of the dwarfs fell in to the Milky Way together, or had past interactions with one another.

\subsubsection{The effect of measurement errors}\label{sec:measure}

In this section, we add the effects of measurement errors. We consider both current typical errors, and those expected in the Gaia era. For current errors, we use $\Delta r=5$\,kpc, $\Delta v_{los}=30$km/s and $\Delta v_{t}=60$km/s, where $\Delta r$ is the distance error and $\Delta v_{los,t}$ are the line of sight and tangential velocity errors, respectively \citep[see e.g.][]{2007AJ....133..818P}. For Gaia errors, we use $\Delta r=5$\,kpc, $\Delta v_{los}=5$km/s and $\Delta v_{t}=10$km/s \citep[see e.g.][]{1999MNRAS.310..645W}. We determine the effects of measurement errors on our derived orbits by building an ensemble of 200 orbits drawn from the above distributions around an already defective orbit, where the defective orbit is also drawn from the same error distribution. This represents the simplest way to mimic the real measurement errors of the Milky Way dwarfs.

\begin{figure*}
\centering
$F$; Current errors\\
\vspace{4.2mm}
\includegraphics[width=0.3\textwidth]{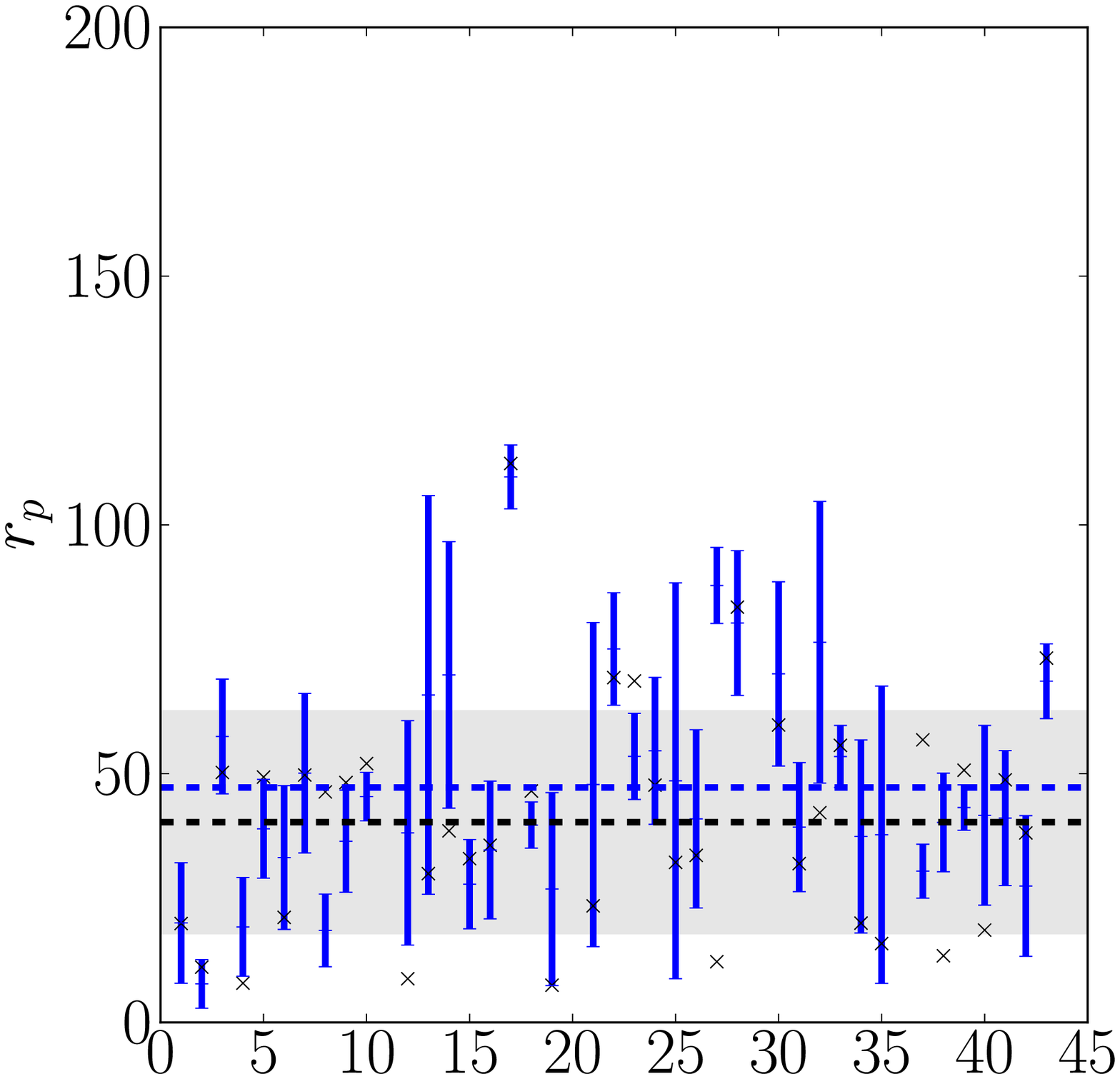}\hspace{3mm}
\includegraphics[width=0.3\textwidth]{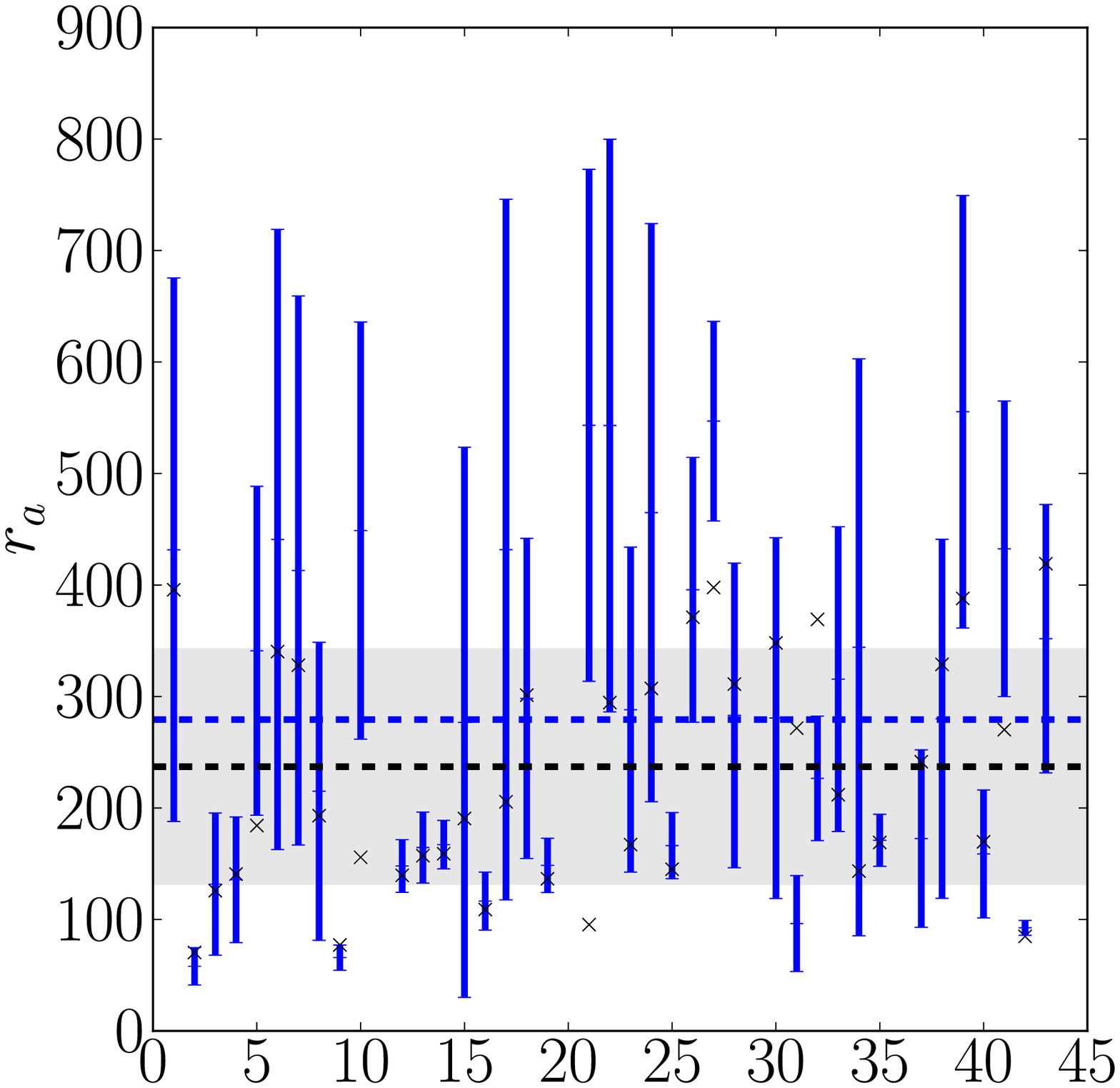}\hspace{3mm}
\includegraphics[width=0.3\textwidth]{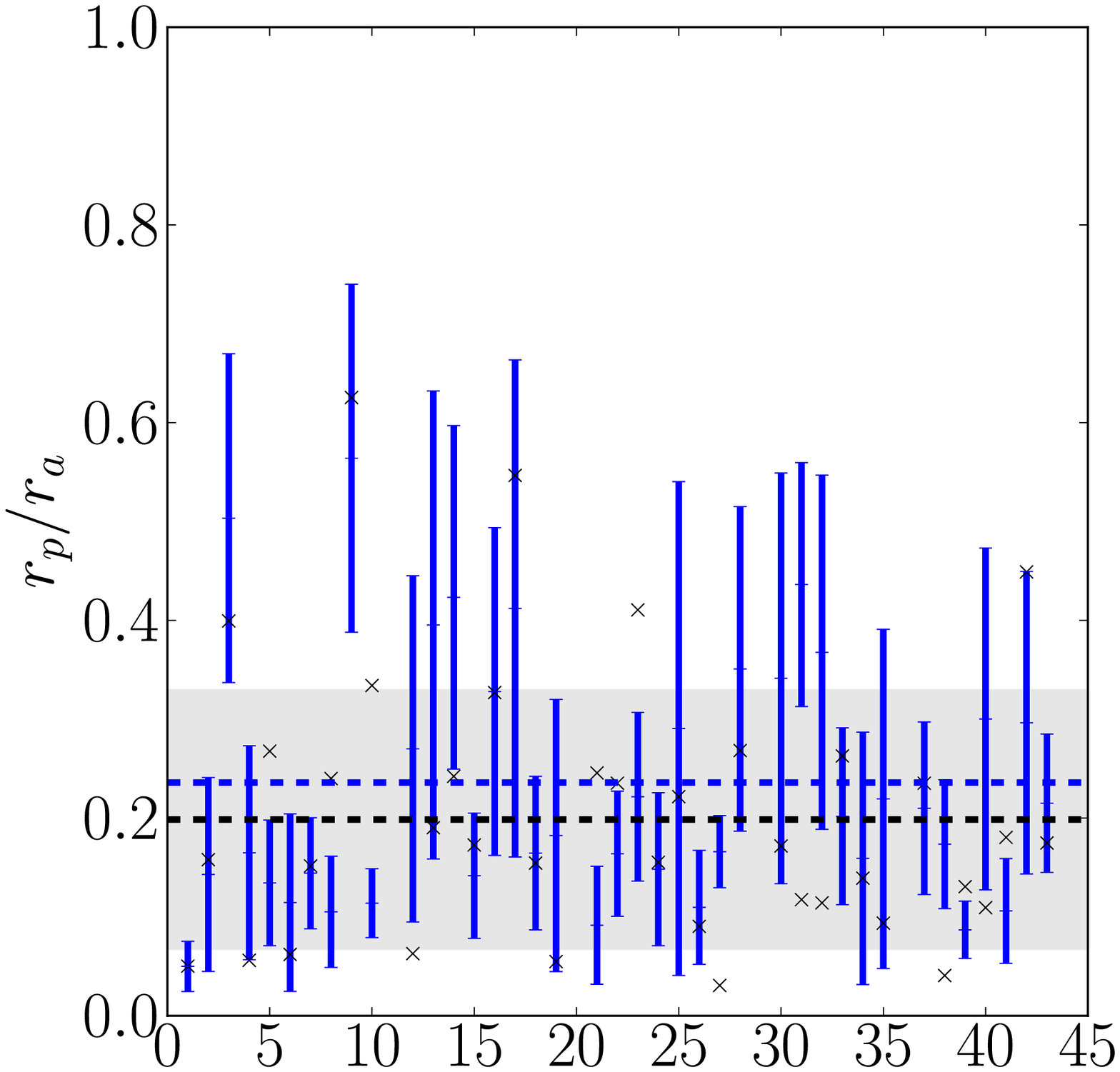}\hspace{3mm}\\
\vspace{5mm}
$F$; Gaia errors\\
\vspace{4.2mm}
\includegraphics[width=0.3\textwidth]{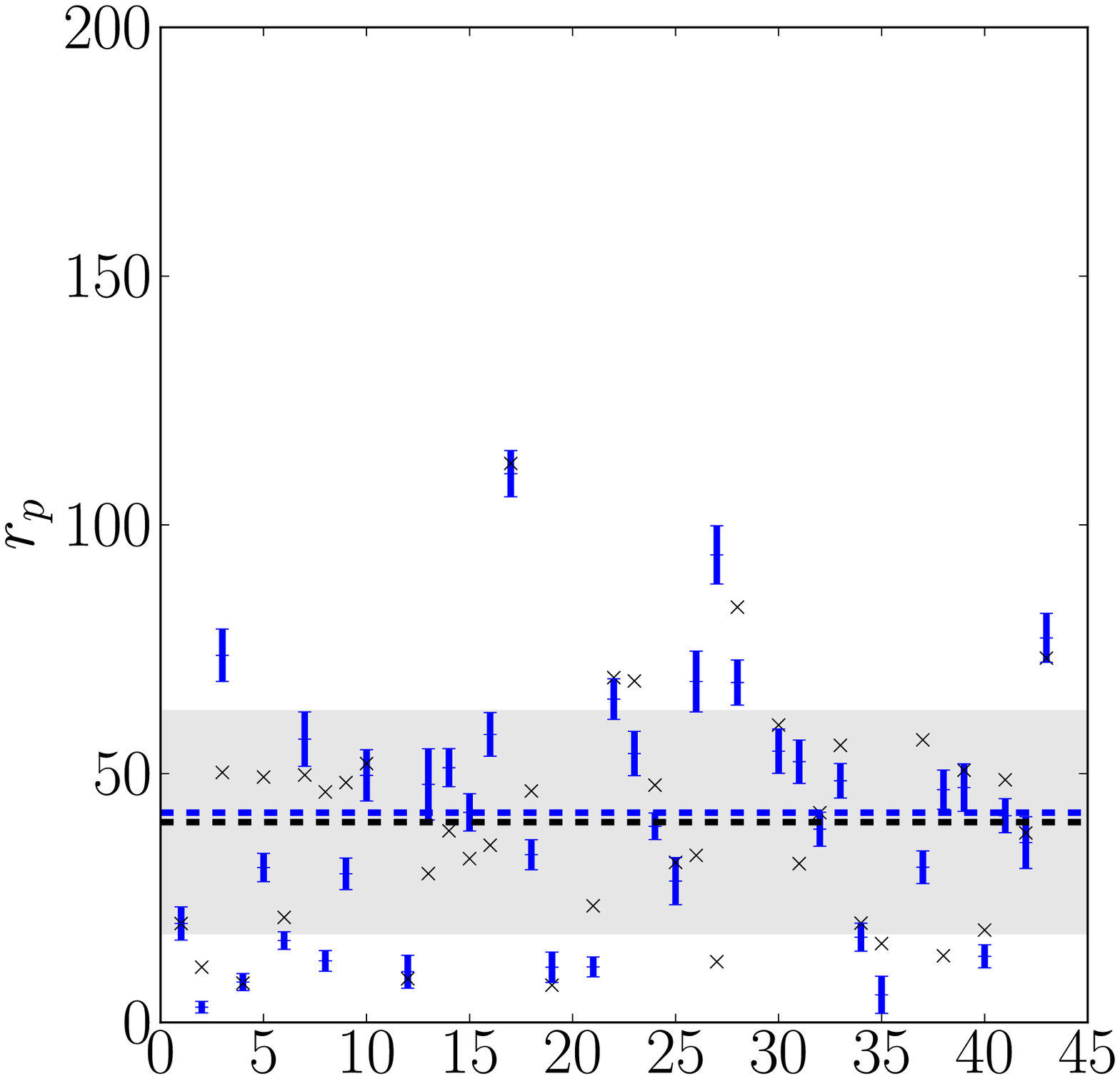}\hspace{3mm}
\includegraphics[width=0.3\textwidth]{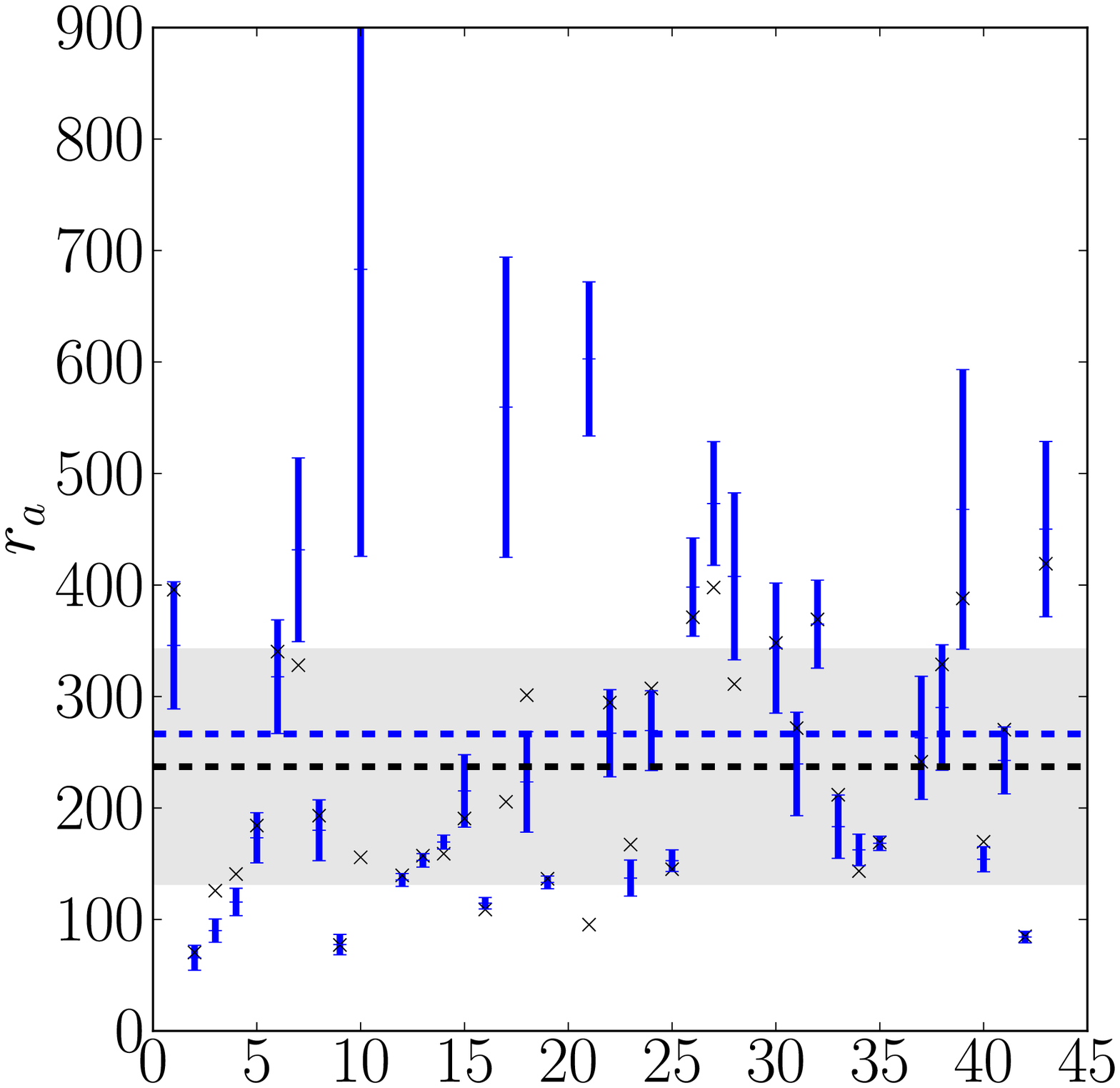}\hspace{3mm}
\includegraphics[width=0.3\textwidth]{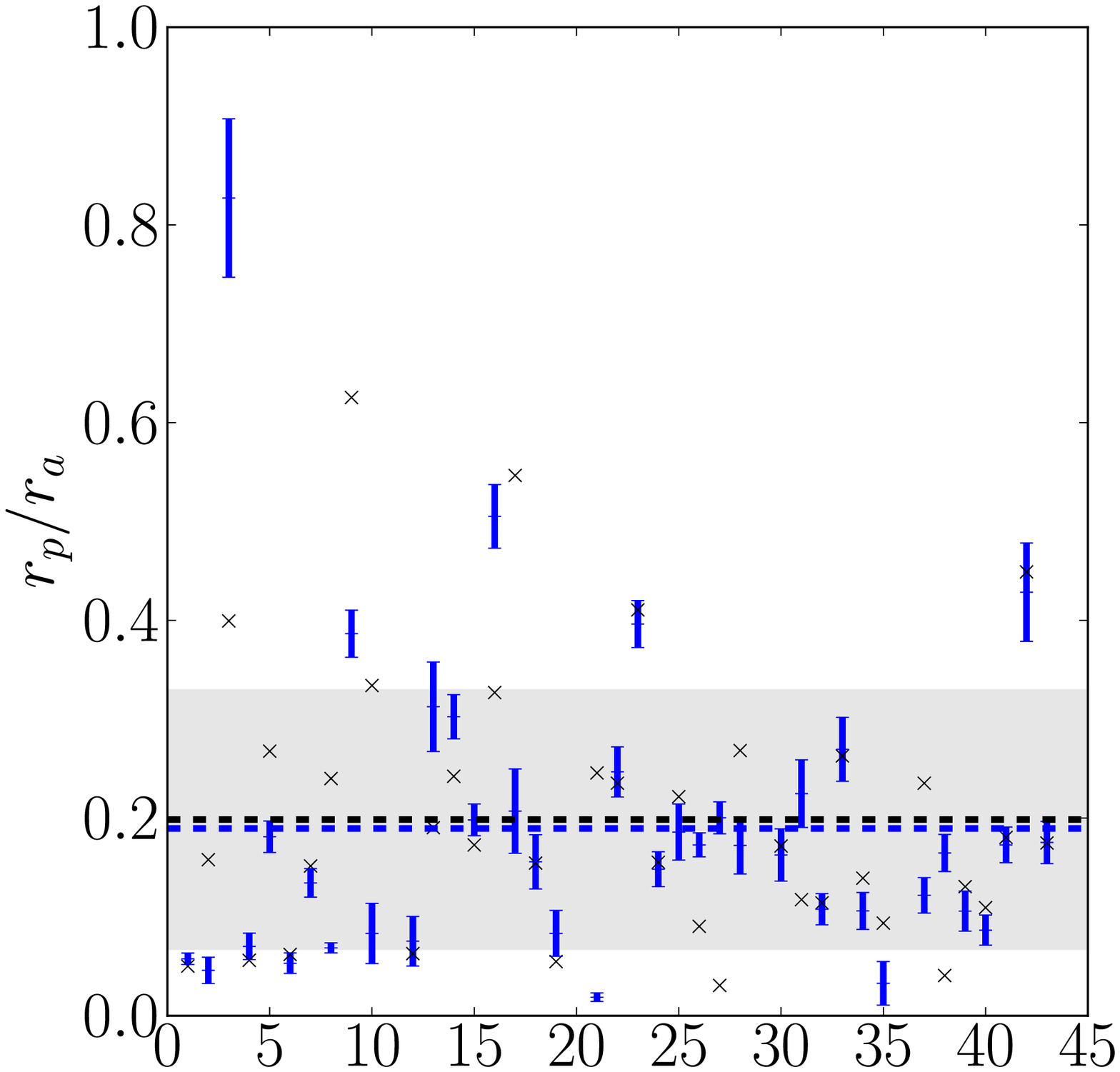}\hspace{3mm}\\
\caption{Orbit recovery in our Fiducial model $F$, with measurement errors. Panels show the pericentre, apocentre and peri/apo ratio for the 50 most massive satellites at $z=0$ in the VL1 simulation. The black crosses correspond to the VL1 data, the blue error bars to the recovered data from the orbit integration. The black dashed horizontal lines and grey shaded area correspond to the mean and the one sigma variation of the true values, respectively. The blue dashed line corresponds to the mean of the recovered values. }
\label{fig:measureF}
\end{figure*}

We show results for our Fiducial model $F$ in Figure \ref{fig:measureF}. We plot the true (black crosses) and recovered (blue error bars) pericentres $r_p$, apocentres $r_a$ and their ratio $r_p/r_a$. The black dashed line shows the mean of the true values, while the grey shaded band marks one sigma scatter around the mean. The blue dashed line is the mean of the recovered orbits.

For current measurement uncertainties, the error is dominated by the proper motion errors ($\Delta v_{t}$). For Gaia errors, we are instead limited by model systematics. However, such systematics appear to average out over the whole population. Although our mean recovered $r_p$ and $r_a$ are both biased high by the proper motion, this biasing is significantly reduced with Gaia quality data. 

\subsubsection{Orbit recovery at levels (i), (ii) and (iii)}\label{sec:orbitrecover}

With current proper motion errors, we are able to recover orbits only at level (i) (the {\it last} pericentre $r_p$ and apocentre $r_a$;  see \S\ref{sec:intro}). With Gaia quality data, it will be possible to recover orbits at level (ii) ($r_p$, $r_a$ and the orbital period $t$ backwards in time over $N \sim 2$ orbits), though the pericentres will suffer from large systematic errors if the Milky Way potential is triaxial. Even with perfect data, going back further than $\sim 2$ orbital periods runs into the problem of group infall (see \S\ref{sec:group}). Full 3D orbit recovery (level (iii)) will be extremely challenging. recovering 3D orbit data will only be possible if the Milky Way potential is nearly axisymmetric or spherical, and if it did not change significantly over the past $\sim 8$\,Gyrs.

%__________________________________________________________
\section{Application to the Milky Way's dwarfs}\label{sec:data}

In this section, we apply our orbit recovery technique to nine Milky Way dwarfs with observed proper motions. We compare our derived orbits to those of subhalos in the VL1 simulation. With current measurement errors, we can only obtain a reliable estimate of the last pericentre and apocentre (see \S\ref{sec:orbitrecover}) and so we focus on these orbit diagnostics.

\subsection{The data}\label{sec:thedata}

Table \ref{tab:dwarfdata1} summarises the dwarf data we take from the literature. From left to right the columns show: galactic longitude $l$ and latitude $b$ in degrees; the distance to the sun $d$ in kpc; the radial velocity $v_r$ in km/s; the proper motion data $\mu_\alpha\cos(\delta)$, $\mu_\delta$ in mas/yr; the V-band magnitude $M_V$\,(mag); and the Mass within 600\,pc in units of $10^7M_\odot$. We order the dwarfs by their star formation history, placing them in three broad categories: those with mainly early star formation ($>9$\,Gyrs; E); those with mainly intermediate age star formation ($3-9$\,Gyrs; I); and those with any significant recent star formation ($<3$\,Gyrs; R). Some half-categories are also introduced for galaxies that have mostly early star formation, but evidence for some intermediate age also (EI). The orbit data recovered in this paper as well as our classification of the star formation histories and the corresponding references are summarised in Table \ref{tab:dwarfdata}. 

\newlength{\myspace}
\setlength{\myspace}{0.013\textwidth}
\begin{table*}
\begin{center}
\setlength{\arrayrulewidth}{0.3mm}
\begin{tabular*}{\textwidth}
{
@{\hspace{\myspace}}c
@{\hspace{\myspace}}c
@{\hspace{\myspace}}c
@{\hspace{\myspace}}c
@{\hspace{\myspace}}c
@{\hspace{\myspace}}c
@{\hspace{\myspace}}c
@{\hspace{\myspace}}c
@{\hspace{\myspace}}c
@{\hspace{\myspace}}c
}
Galaxy & $l$\,(deg) & $b$\,(deg) & $d$\,(kpc) & $v_r$\,(km/s) & $\mu_\alpha\cos(\delta)$\,(mas/yr) & $\mu_\delta$\,(mas/yr) & $M_V$\,(mag) & $M_{600}[10^7M_\odot]$ & Refs.\\ 
\hline
UMi  & 105.0	&+44.8	&$66\pm3$		&$-248\pm2$			&$-0.50\pm0.17$	&$0.22\pm0.16$  &-8.9 &  $5.3^{+1.3}_{-1.3}$ &1,2,12\\ 
Draco  &86.4	&34.7	&$82\pm6$		&$-293\pm2$			&$0.6\pm0.4$		&$1.1\pm0.3$       & -8.8  & $4.9^{+1.4}_{-1.3}$ &1,3,12\\ 
Sextans  &243.5 &+42.3	&$86\pm4$		&$227\pm3$			&$0.26\pm0.41$	&$0.10\pm0.44$   & -9.5 & $0.9^{+0.4}_{-0.3}$ &1,4,12\\ 
Sculptor  &287.5 &-83.2	&$79\pm4$		&$108\pm3$			&$0.09\pm0.13$	&$0.02\pm0.13$  &  -11.1 & $2.7^{+0.4}_{-0.4}$ & 1,5,12\\
Carina  &260.1	&-22.2	&$101\pm5$		&$224\pm3$			&$0.22\pm0.13$	&$0.24\pm0.11$& -9.3  &$3.4^{+0.7}_{-1.0}$ & 1,6,12\\ 
Fornax  &237.1	&-65.7	&$138\pm8$		&$53\pm3$			&$0.476\pm0.046$	&$-0.360\pm0.041$& -13.2  & $4.3^{+2.7}_{-1.1}$ &1,7,12\\ 
Sagittarius  &5.6 &-14.1	&$24\pm2$		&$140\pm5$			&$-2.35\pm0.20^\star$&$2.07\pm0.20^\star$ & -13.4 &  $27^{+20}_{-27}$& 1,8,12\\
SMC  &302.8	&-44.3	&$58$			&$175$				&$1.16\pm 0.18$	&$-1.17\pm0.18$ & -17.2             &  $10\pm2$ &1,9,10,13\\
LMC  &280.5   	&-32.9	&$49$			&$324$				&$2.03\pm0.08$	&$0.44\pm0.05$ & -18.6	            & $14\pm3$ &1,10,11,14\\
\end{tabular*}
\caption[]{Dwarf data from the literature. From left to right, the columns show galactic longitude; galactic latitude; distance to the sun; radial velocity; proper motions; V-band magnitude; the dynamical mass within 600\,pc (M$_{600}$); and the data references. Data were taken from: $^1$\citet{1998ARA&A..36..435M}; $^2$\citet{2005AJ....130...95P}; $^3$\citet{1994IAUS..161..535S}; $^4$\citet{2008ApJ...688L..75W}; $^5$\citet{2006AJ....131.1445P}; $^6$\citet{2003AJ....126.2346P,2004AJ....128..951P}; $^7$\citet{2007AJ....133..818P}; $^8$ \citet{2005ApJ...618L..25D}; $^9$ \citet{2006ApJ...652.1213K}; $^{10}$ \citet{2009ApJ...696.2179K} ; $^{11}$ \citet{2006ApJ...638..772K}; $^{12}$ \citet{2007ApJ...669..676S}; $^{13}$ mass derived from mean of \citet{1997A&AS..122..507H,2004ApJ...604..176S, 2006AJ....131.2514H,2008MNRAS.386..826E}; $^{14}$ mass derived from \citet{1998ApJ...503..674K}, consistent with \citet{2002AJ....124.2639V}; $^\star$ these values are for $\mu_l\cos(b)$ and $\mu_b$, respectively.}
\label{tab:dwarfdata1}
\end{center}
\end{table*}

\begin{table}
\begin{center}
\setlength{\arrayrulewidth}{0.3mm}
\begin{tabular*}{0.5\textwidth}
{
@{\hspace{\myspace}}c
@{\hspace{\myspace}}c
@{\hspace{\myspace}}c
@{\hspace{\myspace}}c
@{\hspace{\myspace}}c
@{\hspace{\myspace}}c
}
Galaxy & $r_p$\,(kpc) & $r_a$\,(kpc) & $T$\,(Gyrs) & SFH & Refs. \\
\hline
UMi (L05) & $40\pm20$	& $90\pm20$ & $1.4\pm0.4$& E & 1,2\\
UMi  (TF)& $30\pm10$ 	& $80\pm10$ & $1.2\pm0.2$ & E & 1,2\\
Draco (L05) & $90\pm 10$ 	& $300\pm100$ & $6\pm2$ & E & 1,3\\
Draco (TF) & $74\pm6$		& $250\pm80$	& $4\pm2$	& E & 1,3\\
Sextans (L05) & $70\pm 20$ & $300\pm200$ & $4\pm4$ & E & 1,4\\
Sextans (TF) & $60\pm 20$ & $200\pm100$ & $4\pm3$ & E & 1,4\\
Sculptor  (L05)& $60\pm10$&  $160\pm80$ & $3\pm1$ & EI & 1\\
Sculptor  (TF)& $60\pm10$ & $130\pm60$ & $2\pm1$ & EI & 1\\
Carina (L05) & $60\pm 30$ &  $110\pm30$& $2.0\pm0.6$ & I & 1\\
Carina (TF) & $50\pm 30$ & $110\pm30$& $1.8\pm0.8$ & I & 1\\
Fornax (L05)  &$120\pm 20$ & $180\pm50$& $4\pm1$ & I & 1,5\\
Fornax (TF) & $110\pm20$ & $170\pm40$ & $4\pm1$ & I & 1,5\\
Sagittarius  (L05)&$12\pm1$ & $50\pm7$& $0.7\pm0.1$ & I & 1\\
Sagittarius  (TF)&$12\pm1$  & $53\pm5$ & $0.56\pm0.08$ & I & 1\\
SMC (L05) & $57\pm5$ 	& $200\pm100$ & $4\pm1$ & R & 1,6\\
SMC (TF) & $56\pm5$ & $200\pm100$ & $3\pm2$ & R & 1,6\\
LMC (L05) & $47\pm1$& $500\pm100$ & $7\pm2$ & R & 1,7\\
LMC (TF) & $47\pm1$ & $400\pm100$ & $7\pm3$& R & 1,7\\
\end{tabular*}
\caption[]{Derived orbits for the dwarfs, along with SFHs from the literature. From left to right, the columns show: Galaxy name; pericentre in kpc; apocentre in kpc; period in Gyrs; the Star formation history classification (see \S\ref{sec:thedata}); and the data references. Each galaxy appears twice to show results from the L05 potential and the TF potential (see \S\ref{sec:potentials}). Notice, that even though the apo- and pericentre are biased high with current measurement errors, the period is biased low. This is due to the bias towards more circular orbits. The star formation history data were taken from: $^1$\citet{2005astro.ph..6430D}; $^2$\citet{2002AJ....123.3199C}; $^3$\citet{2001AJ....122.2524A}; $^4$\citet{2009ApJ...703..692L}; $^5$\citet{2008ApJ...685..933C}; $^6$\citet{2009ApJ...705.1260N}; $^7$\citet{2009AJ....138.1243H}.}
\label{tab:dwarfdata}
\end{center}
\end{table}

\subsection{Potentials}\label{sec:potentials}

For the Milky Way potential, we use two different mass models from the literature. This gives us a handle on the systematic error arising from our potential model. For our first model, we use the oblate potential from \cite{2005ApJ...619..807L} consisting of a Miyamoto-Nagai disc \citep{1975PASJ...27..533M}, Hernquist spheroid \citep{1990ApJ...356..359H}, and a logarithmic halo. The L05 model: 

\begin{equation}
        \Phi_{\rm disc}=- {GM_{\rm disc} \over
                 \sqrt{R^{2}+(a+\sqrt{z^{2}+b^{2}})^{2}}},
\end{equation}
\begin{equation}
        \Phi_{\rm sphere}=-{GM_{\rm sphere} \over r+c},
\end{equation}
\begin{equation}
        \Phi_{\rm halo}=v_{\rm halo}^2 \ln (R^{2}+(z^2/q^2)+d^{2}).
\end{equation}
with $M_{\rm disc}=1.0 \times10^{11} M_{\odot}$; $a=6.5$\,kpc; $b=0.26$\,kpc; $M_{\rm sphere}=3.4 \times 10^{10} M_{\odot}$; $c=0.7$\,kpc; $v_{\rm halo} = 121$\,km/s; and $q=0.9$.

For our second model, we use the Truncated Flat model from \cite{1999MNRAS.310..645W}. The TF model: 

\begin{equation}
\Phi(r) = {GM\over a}\log\Bigl({\sqrt{r^2 + a^2} + a \over r}\Bigr).
\end{equation}
with values  $M = 1.9 \times 10^{12} M_\odot$; and $a = 170$\,kpc. 

These different models have quite different asymptotic total masses, but the mass within the orbit of the dwarfs ($\sim 150$\,kpc) is in reasonable accord ($\sim 6\times10^{11}M_\odot$ in the L05 model, and $12\times10^{11}M_\odot$ in the TF model). The value for the VL1 is $10\times10^{11} M_\odot$ and well within that range. We can therefore directly compare our results without the need to rescale the simulation.

\subsection{Results \& discussion}\label{sec:finalresults} 

\begin{figure*}
\centering
\includegraphics[width=0.4\textwidth]{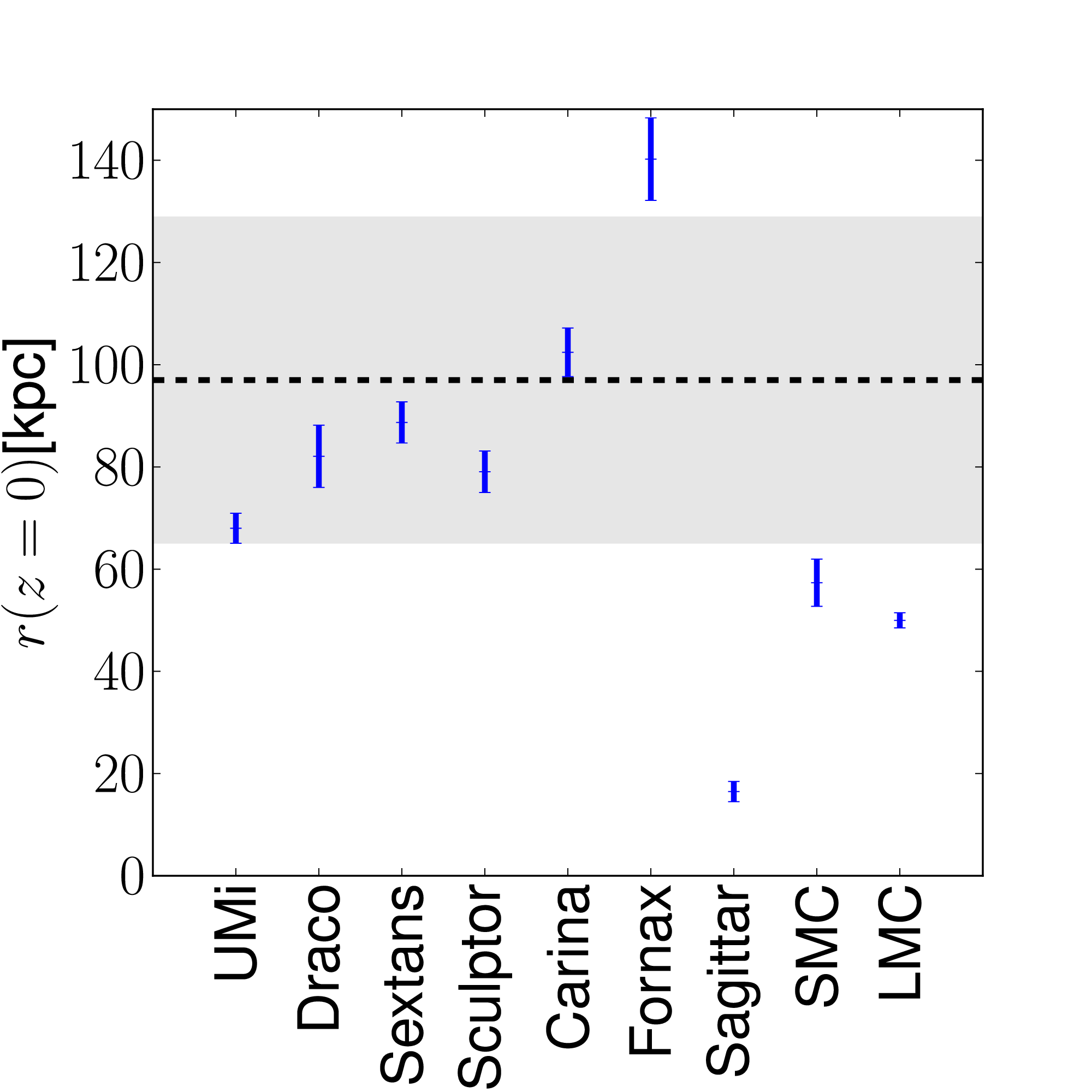}
\includegraphics[width=0.4\textwidth]{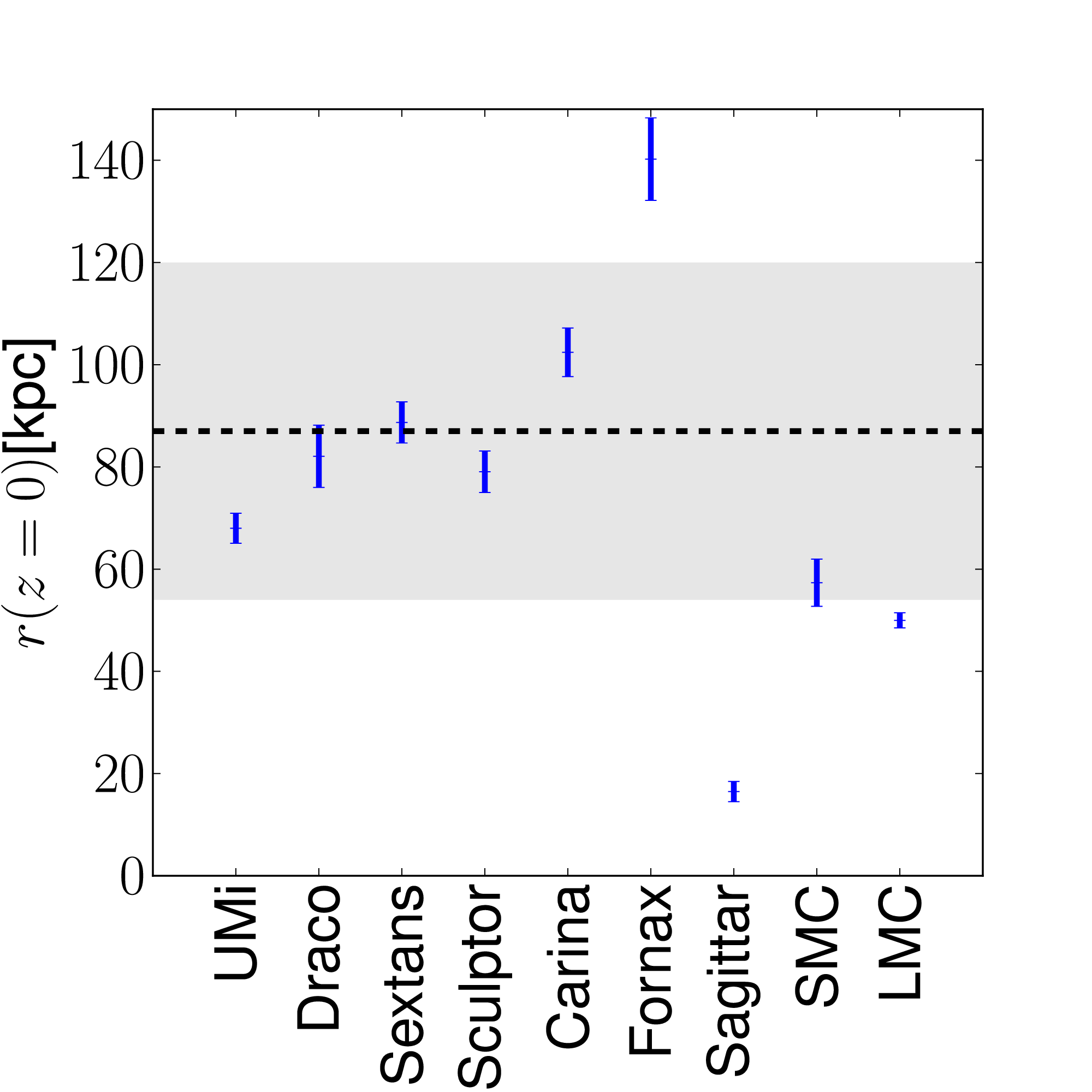}
\caption{Current radii for nine Milky Way dwarfs with observed proper motions with errors. Overlaid are the mean and standard deviation of the values from VL1 for the 50 most massive at $z=0$ (left panel) and for the 50 most massive before $z=10$ (right panel), respectively (grey band, dashed line).}
\label{fig:radial}
\end{figure*}

\begin{figure*}
\centering
\includegraphics[width=0.4\textwidth]{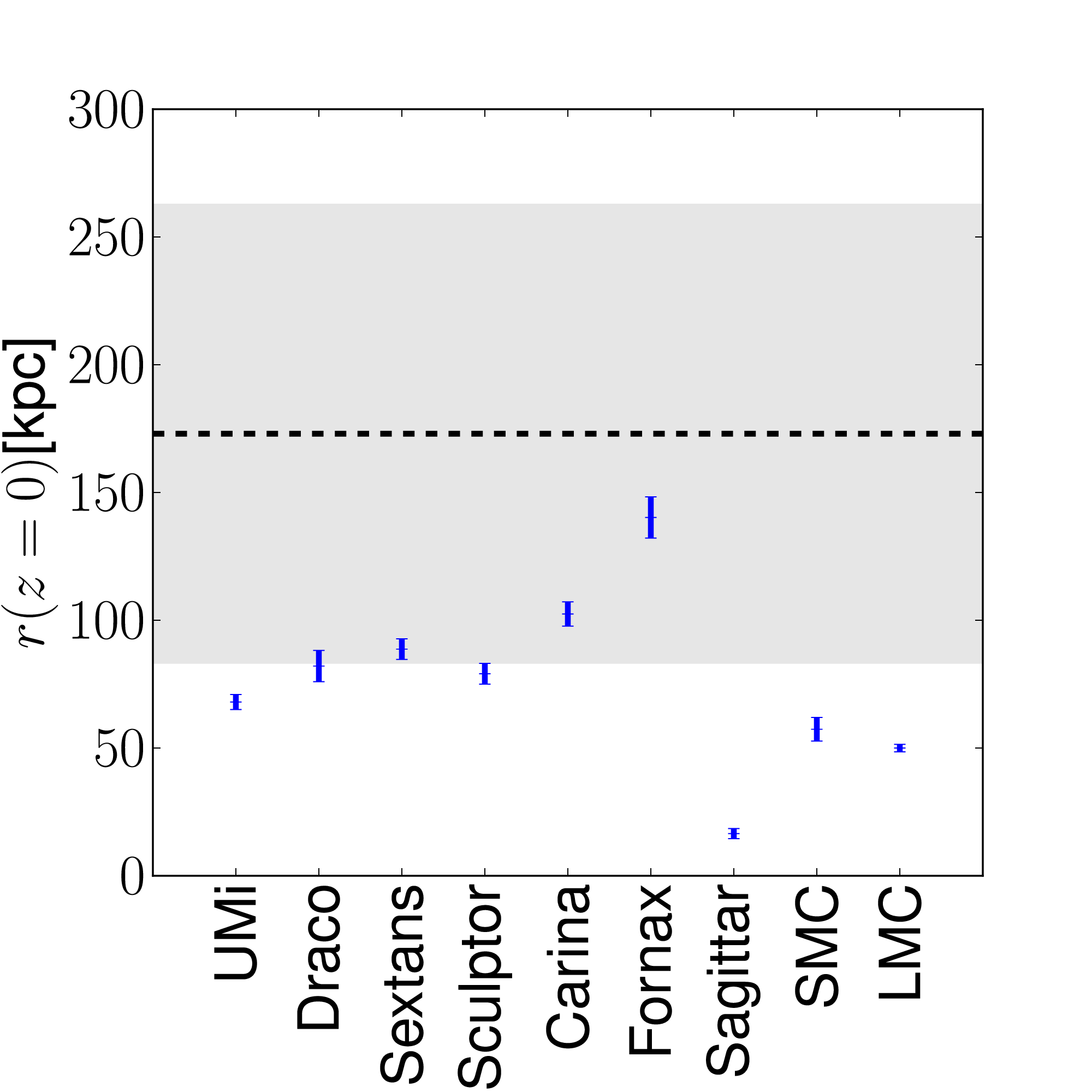}
\includegraphics[width=0.4\textwidth]{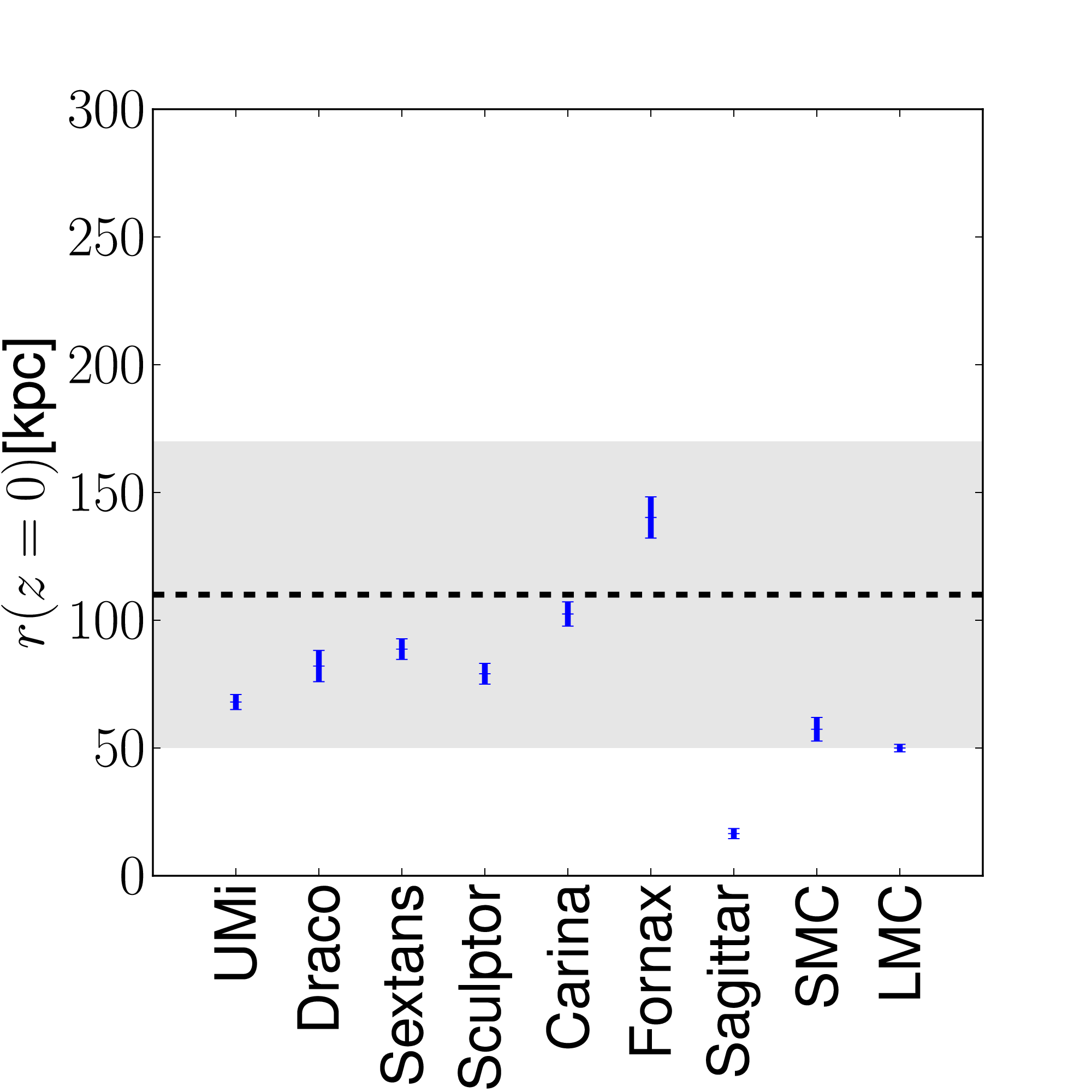}
\caption{As Figure \ref{fig:radial}, but for the radial distributions averaged over the last 2\,Gyrs.}
\label{fig:radialav}
\end{figure*}

\begin{figure*}
\centering
$z^{50}_0$ sample\\
\includegraphics[width=0.3\textwidth]{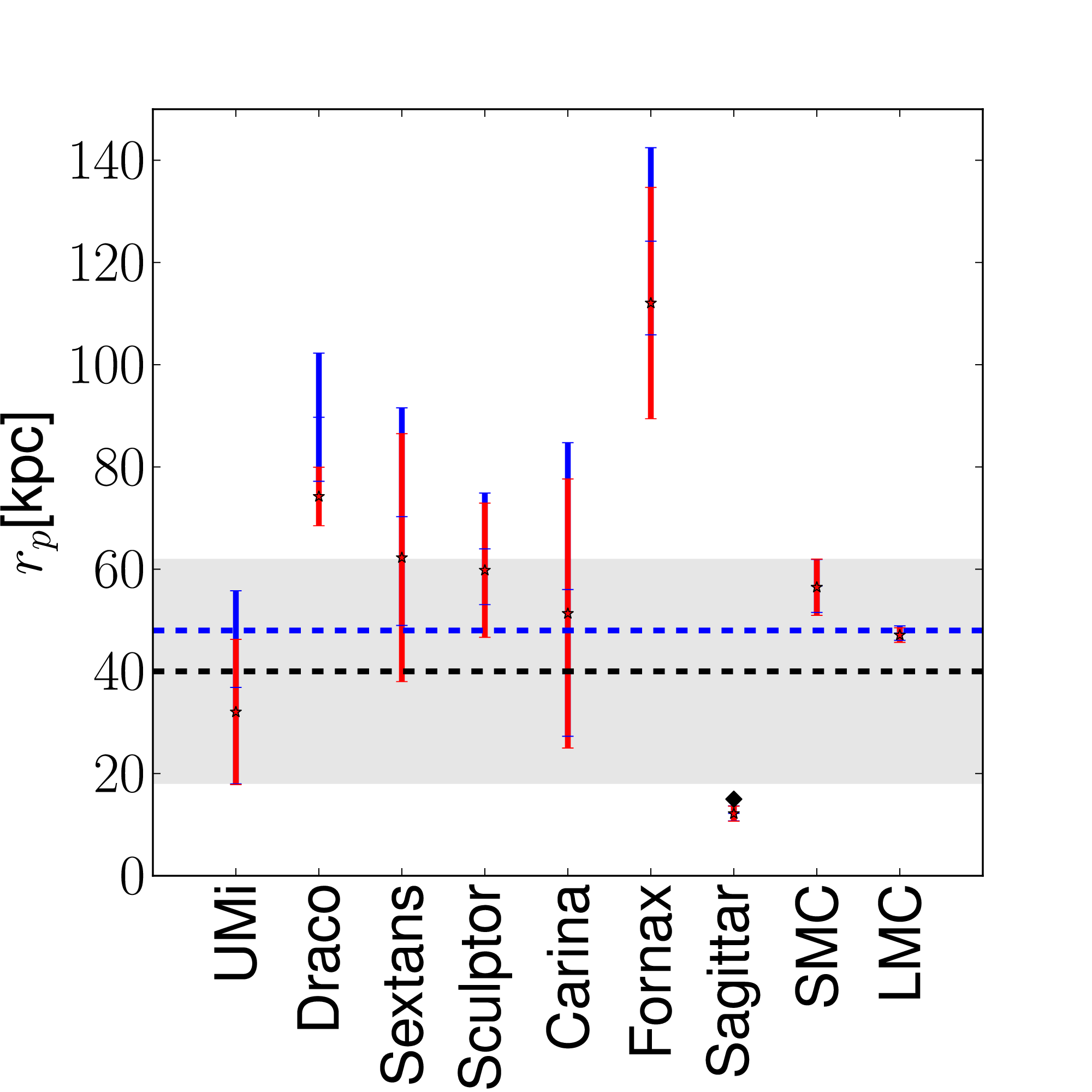}
\includegraphics[width=0.3\textwidth]{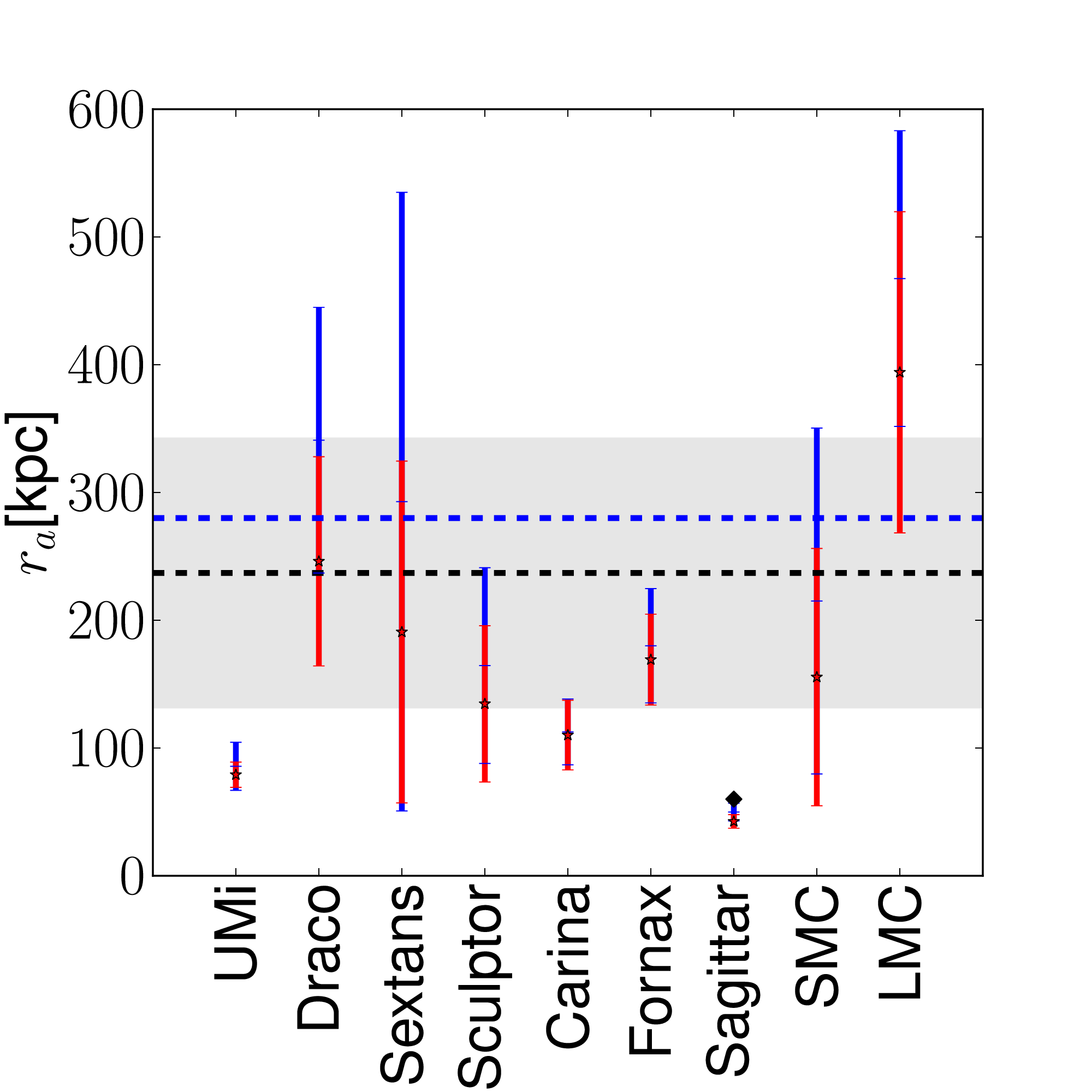}
\includegraphics[width=0.3\textwidth]{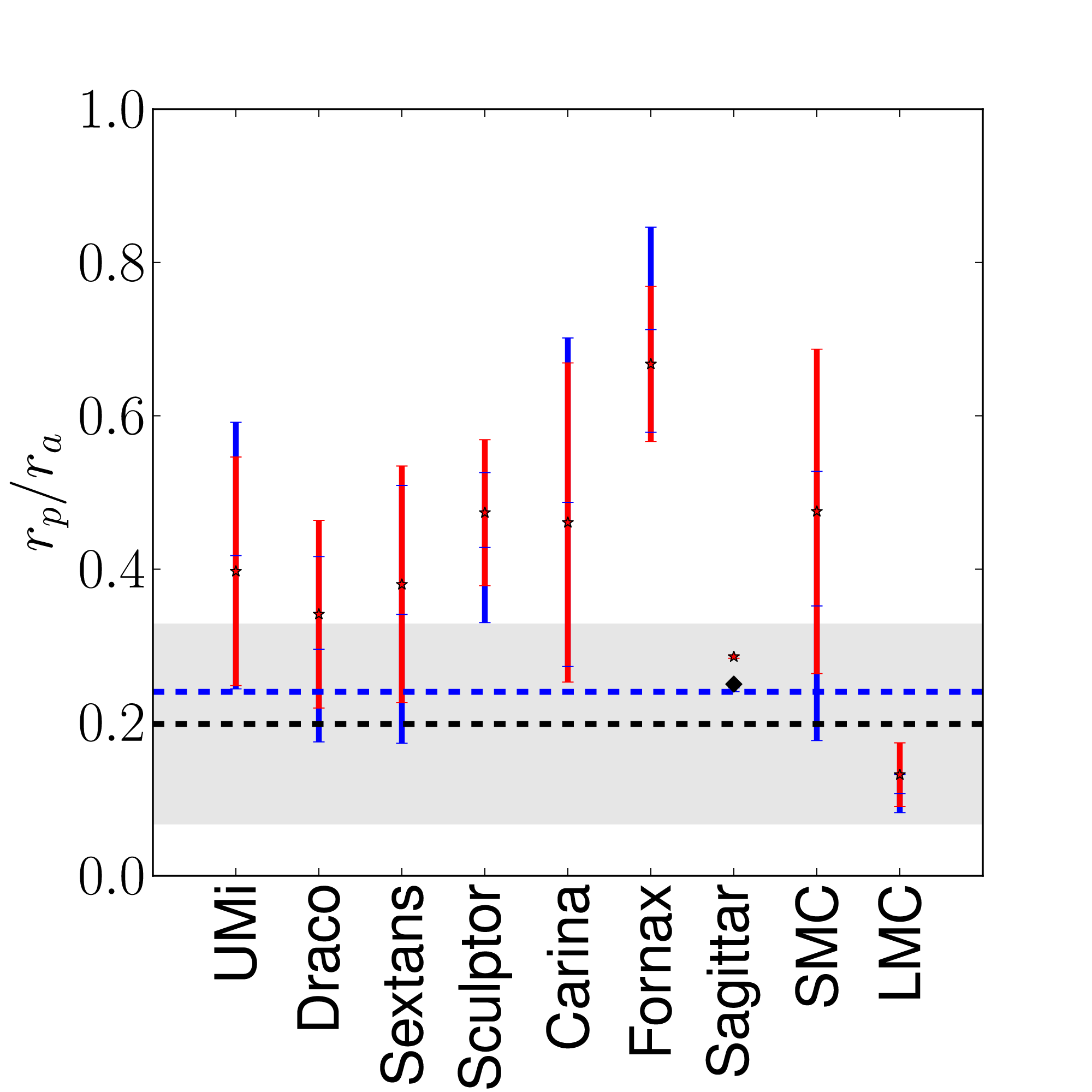}\\
\vspace{10 mm}
$z^{50}_{10}$ sample\\
\includegraphics[width=0.3\textwidth]{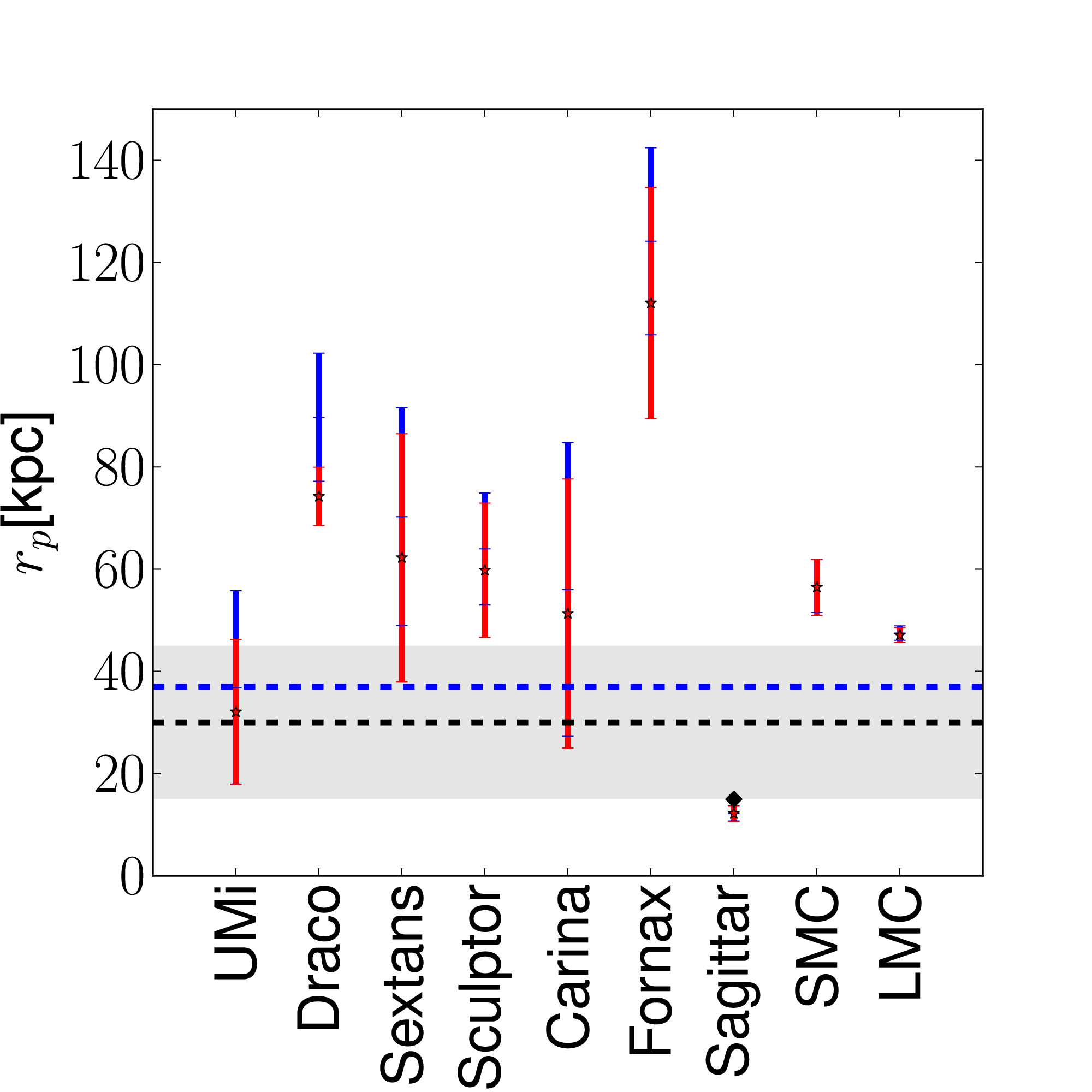}
\includegraphics[width=0.3\textwidth]{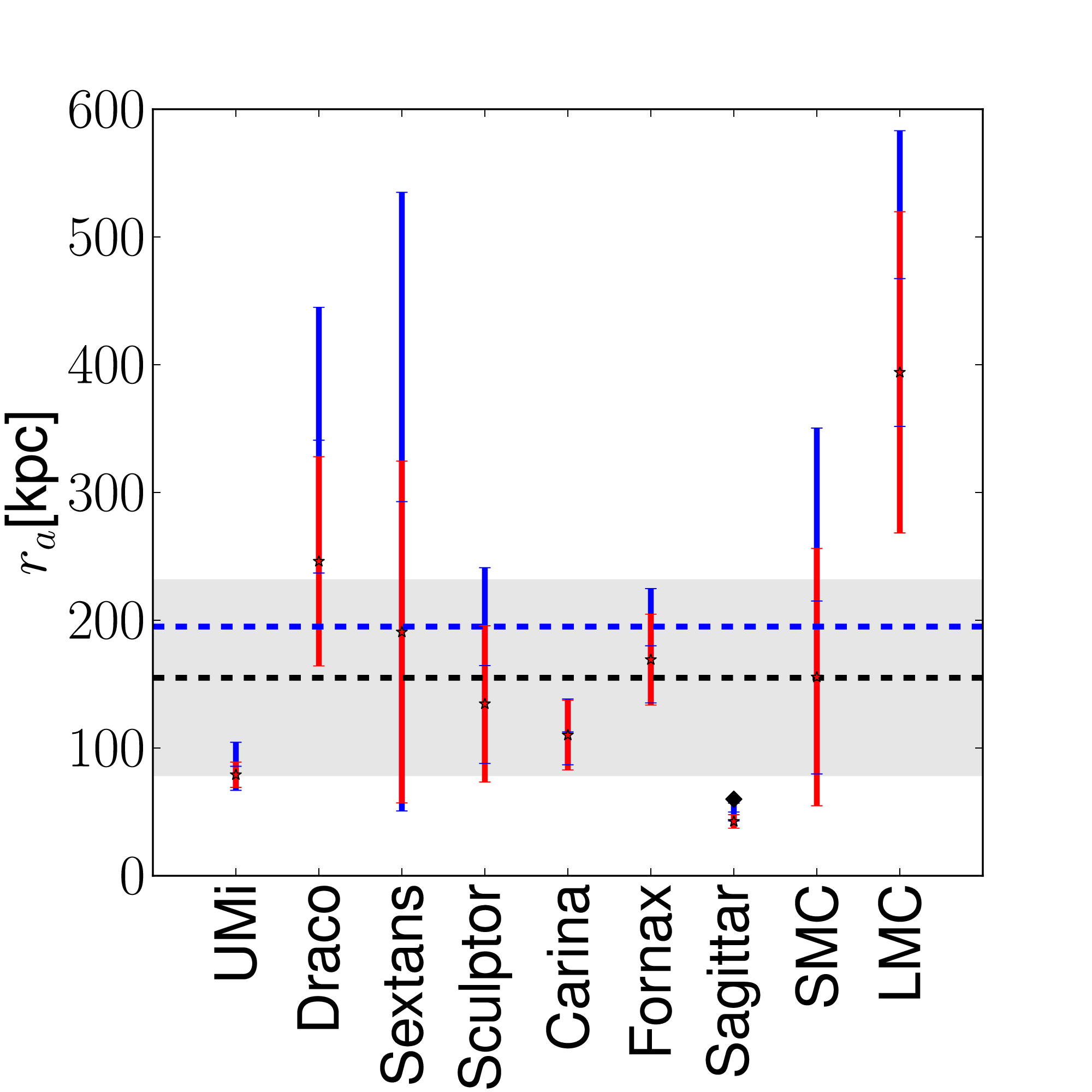}
\includegraphics[width=0.3\textwidth]{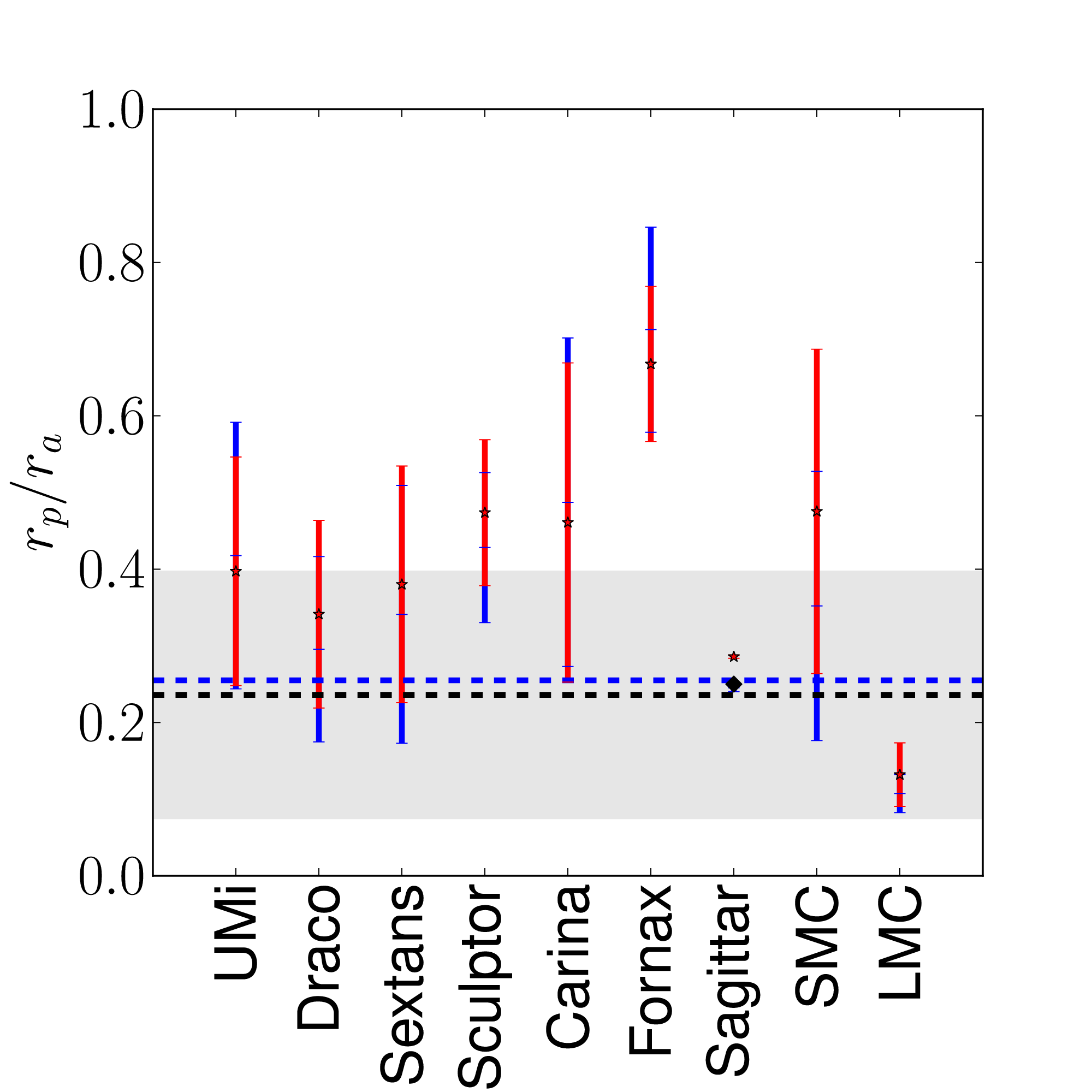}

\caption{Recovered orbits for nine Milky Way dwarfs with observed proper motions. The error bars show the values for the oblate L05 potential (blue) and for the TF model (red). Overlaid are the mean and standard deviation of the values from VL1 for the 50 most massive at $z=0$ and for the 50 most massive before $z=10$, respectively (grey band, dashed line). We also overlay the recovered mean as in Figure \ref{fig:measureF} (blue dashed line). Note that these lines indicate a tendency, as they only represent one possible realisation of the satellite distribution. The galaxies are ordered by their observed star formation histories as described in \S\ref{sec:thedata}. The black diamonds denote the values derived from the Sagittarius stream \citep{2005ApJ...619..807L}.}
\label{fig:dwarfs}
\end{figure*}

\begin{figure*}
\centering
$z^{50}_{10}(r_d=10$\,kpc) sample\\
\includegraphics[width=0.3\textwidth]{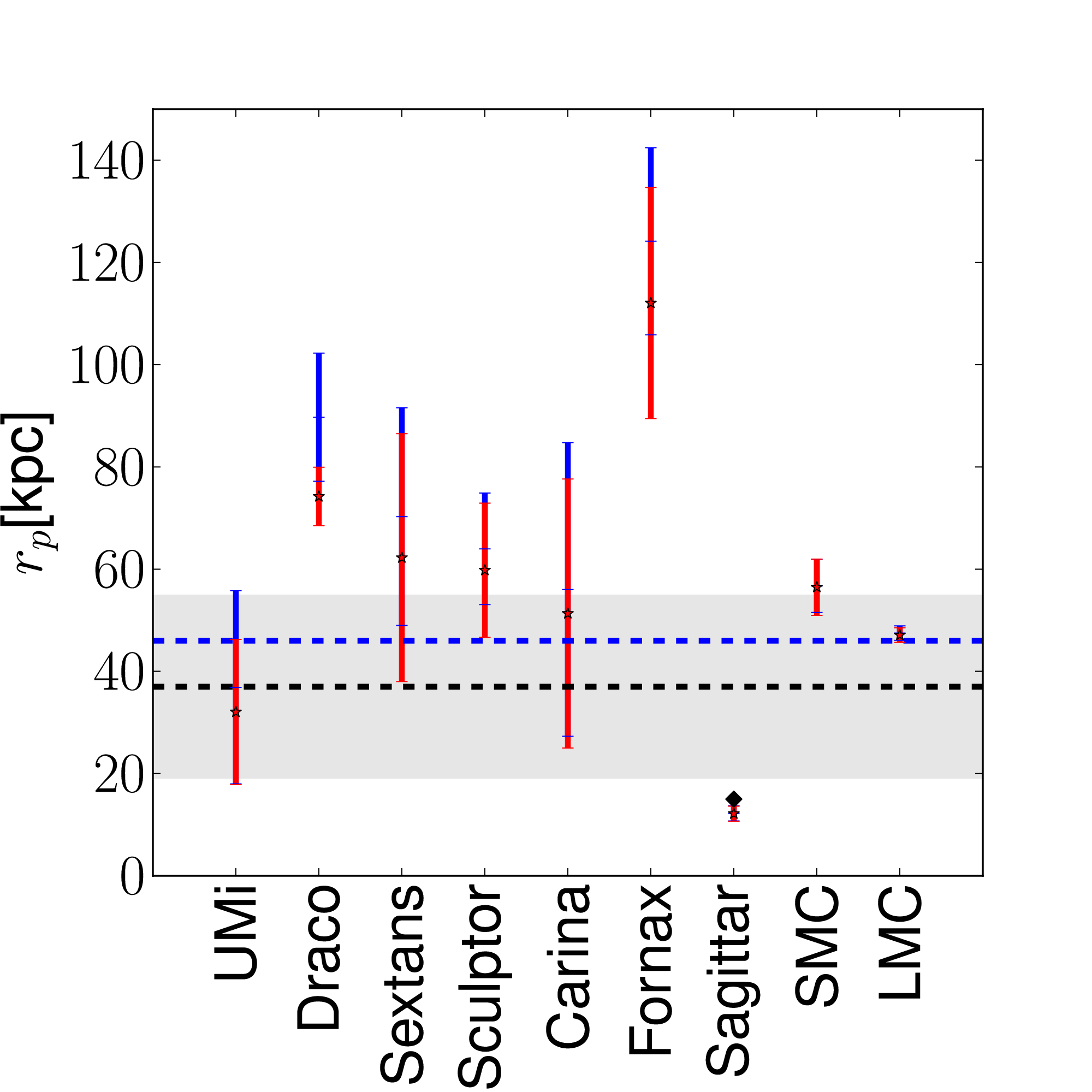}
\includegraphics[width=0.3\textwidth]{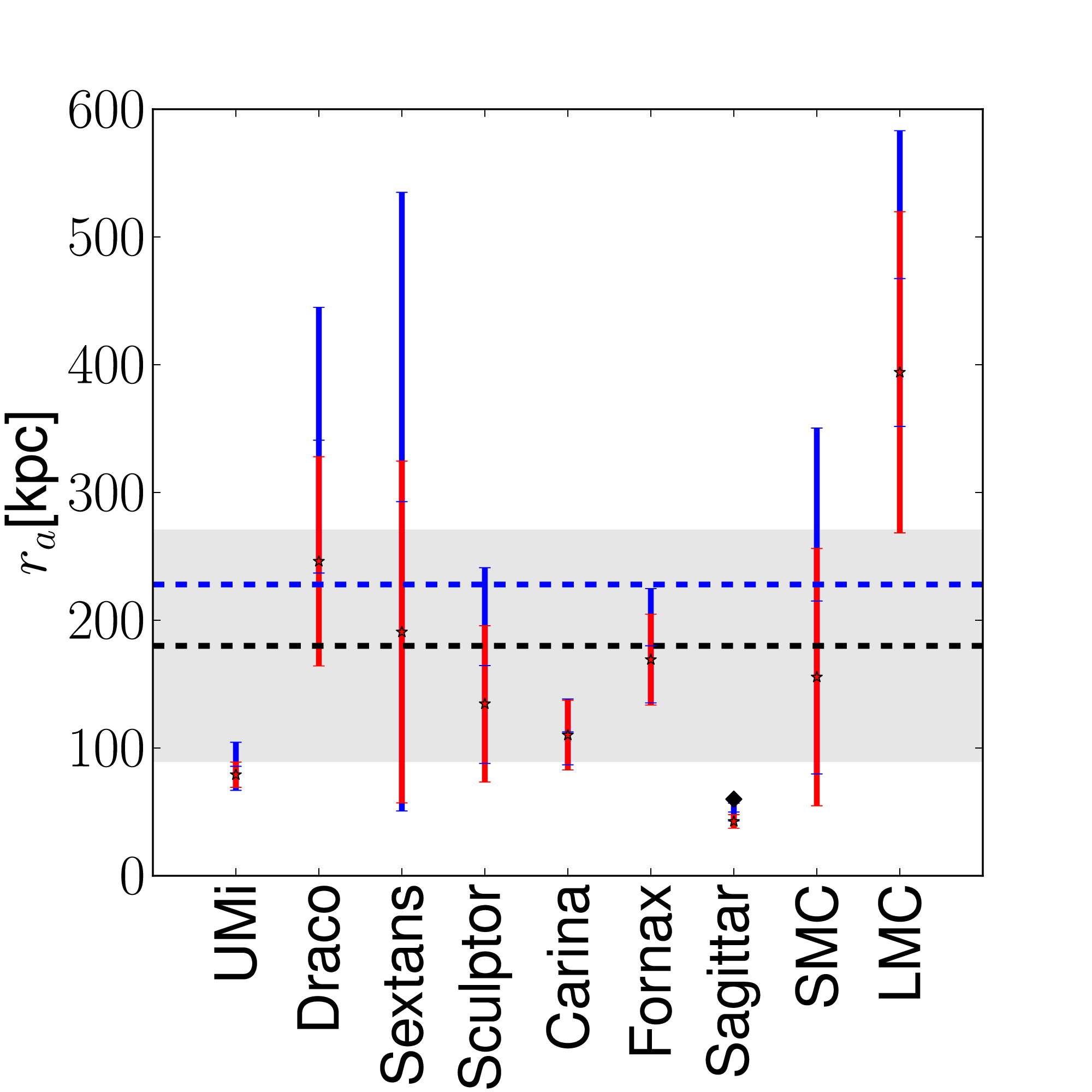}
\includegraphics[width=0.3\textwidth]{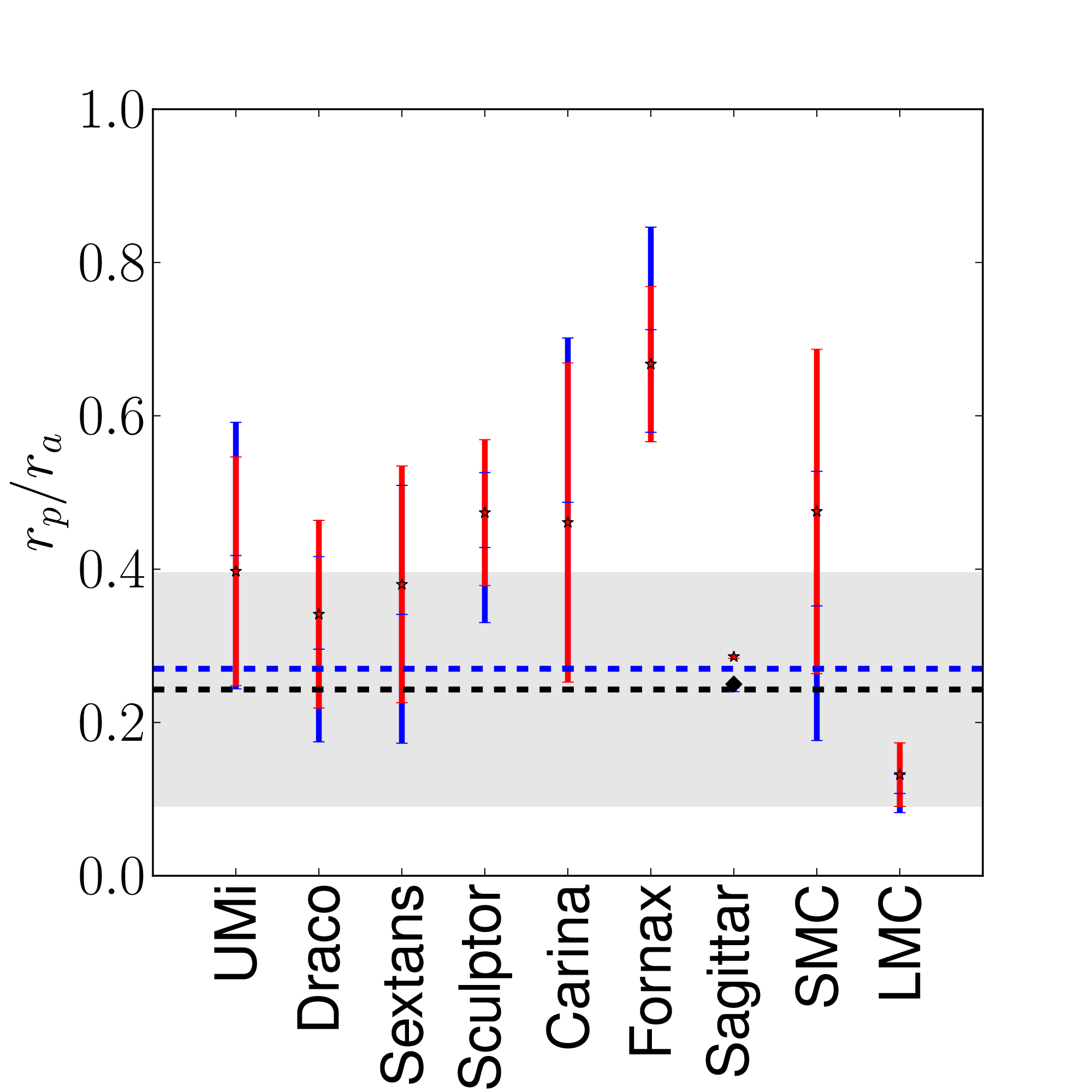}\\
\vspace{10 mm}
$z^{50}_{10}(r_d=15$\,kpc) sample\\
\includegraphics[width=0.3\textwidth]{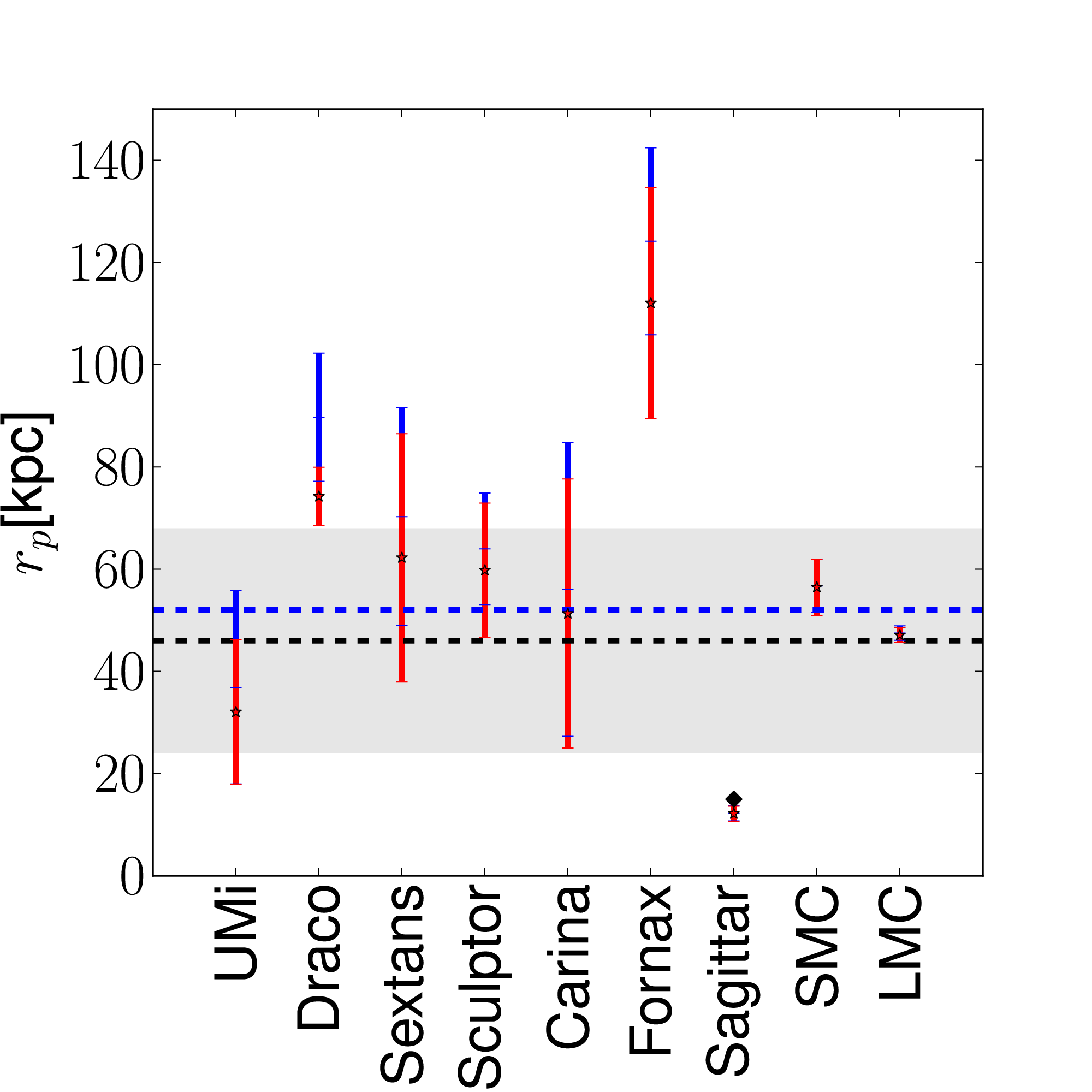}
\includegraphics[width=0.3\textwidth]{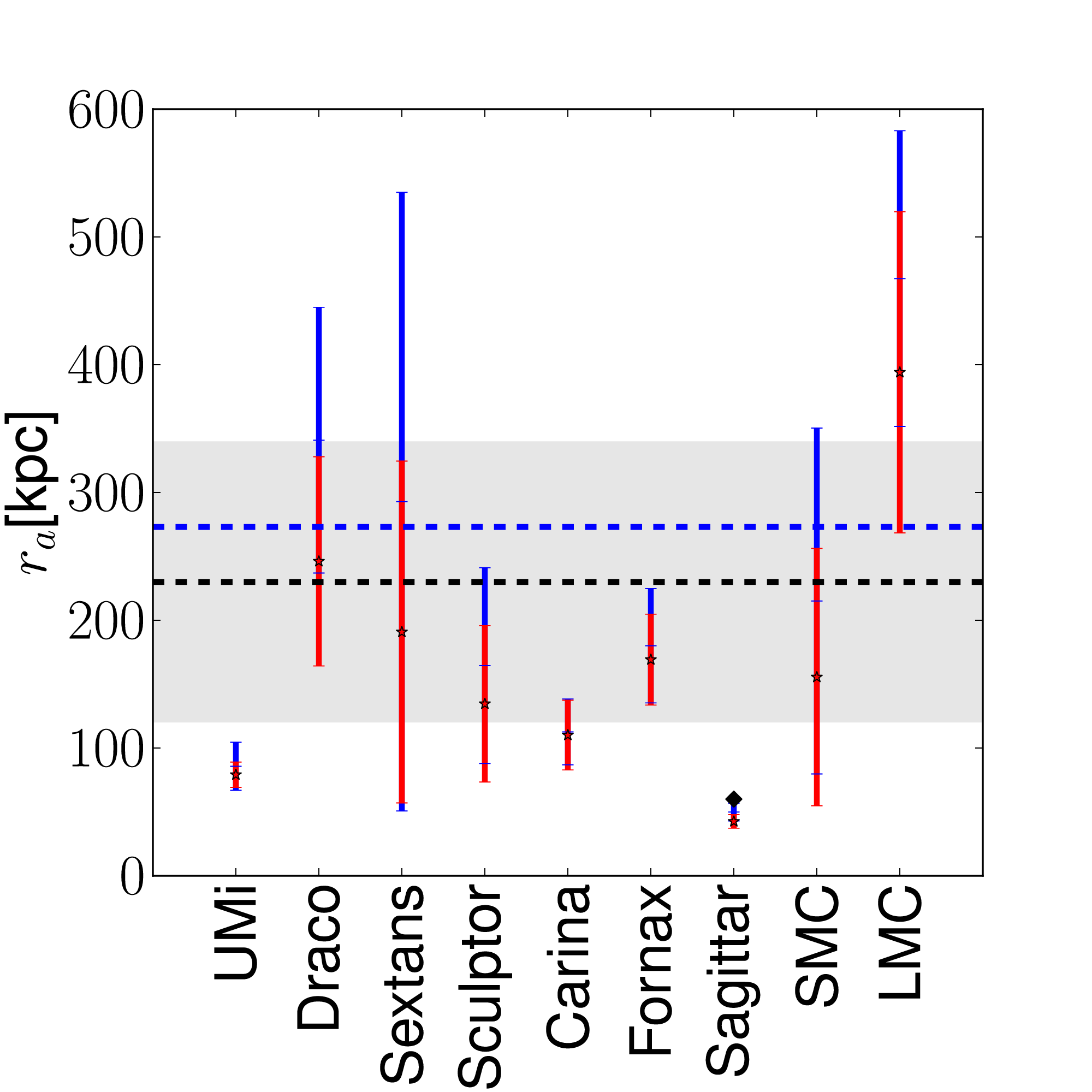}
\includegraphics[width=0.3\textwidth]{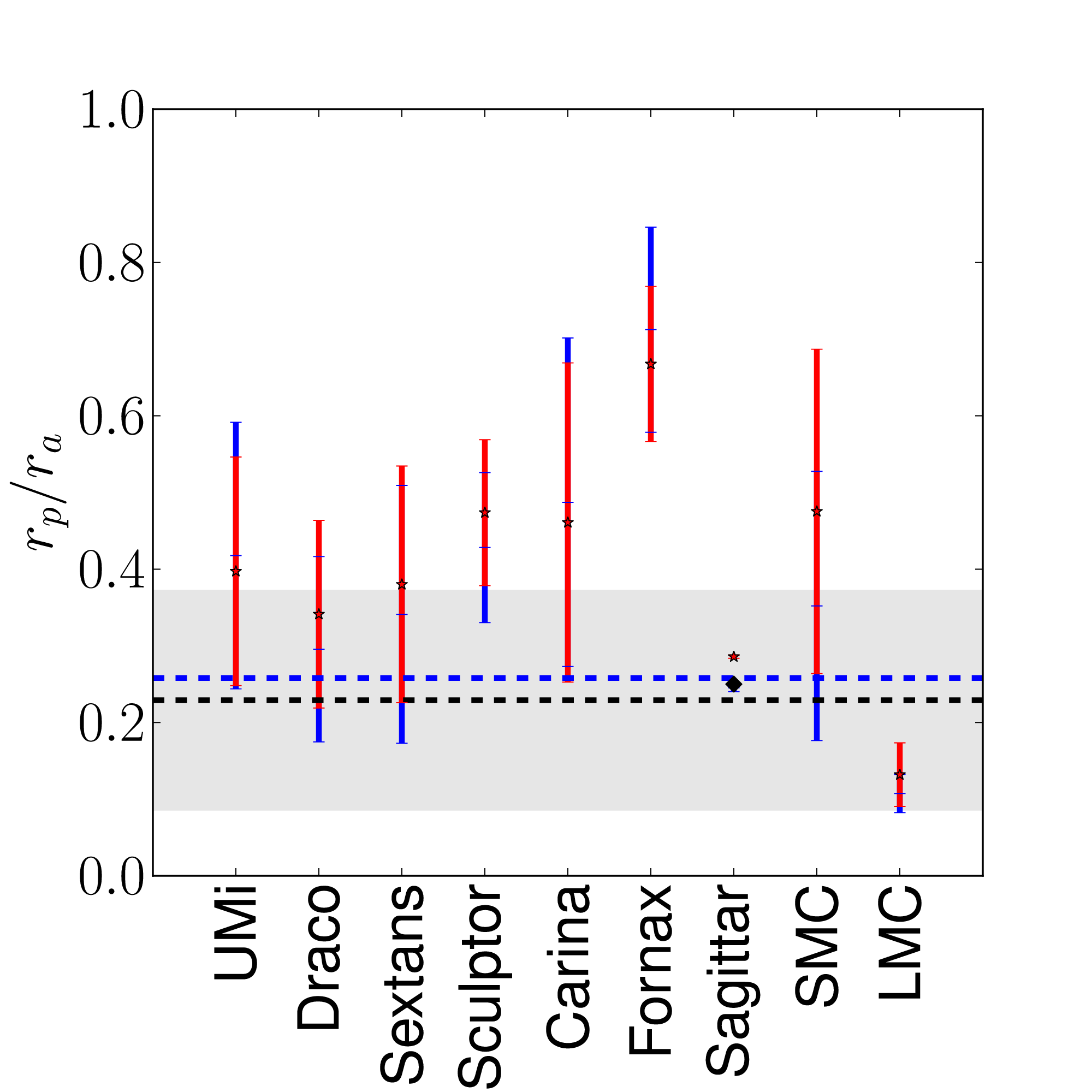}\\
\vspace{10 mm}
$z^{50}_{10}(r_d=20$\,kpc) sample\\ 
\includegraphics[width=0.3\textwidth]{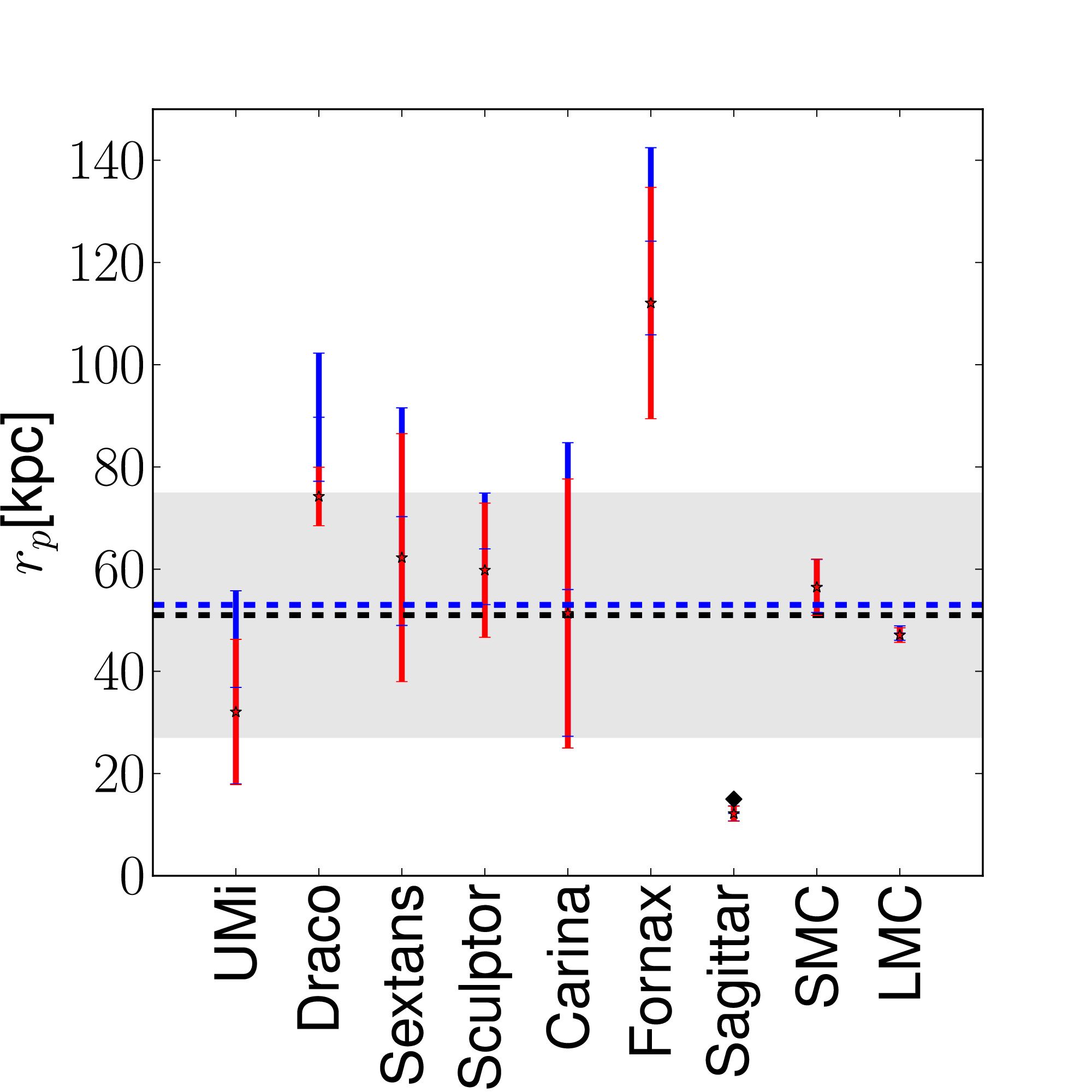}
\includegraphics[width=0.3\textwidth]{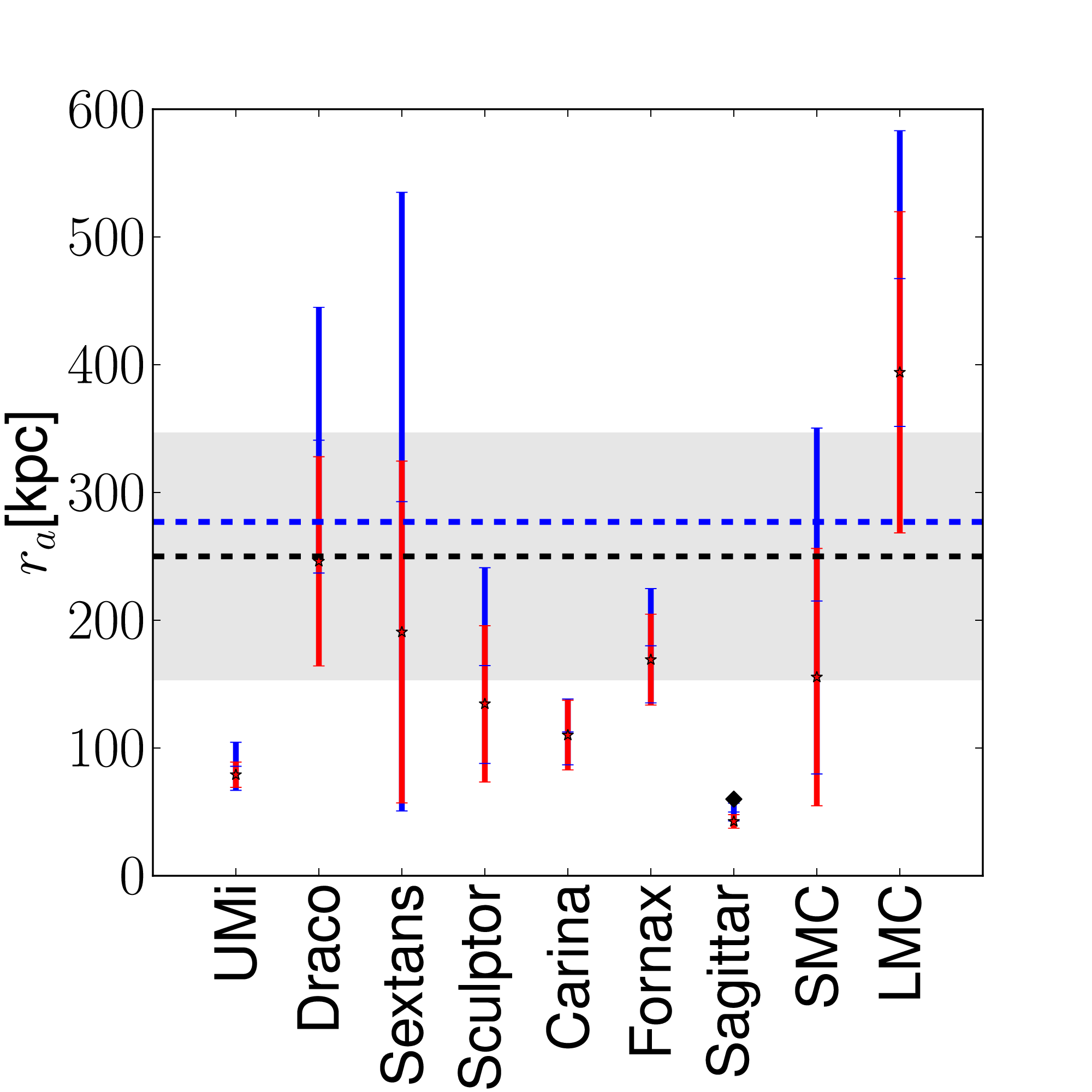}
\includegraphics[width=0.3\textwidth]{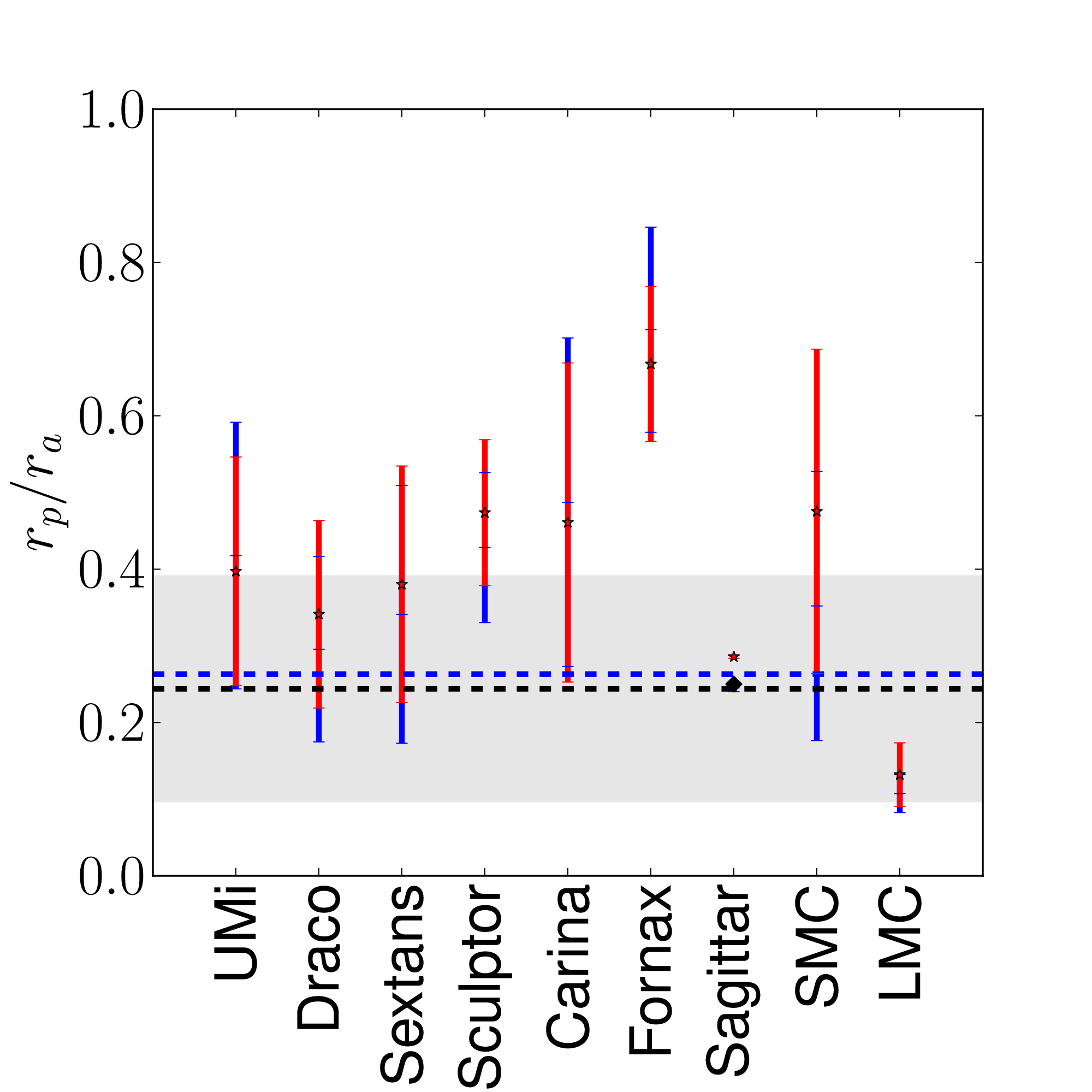}
\caption{As in Figure \ref{fig:dwarfs}, but for three different samples taking disc depletion into account.}
\label{fig:dwarfs2}
\end{figure*}

In Figure \ref{fig:radial}, the current radial distribution of the nine MW dwarfs with errors are shown. Overlaid are the current radial distributions of the two samples $z^{50}_0$ and $z^{50}_{10}$. In seeming contradiction to former results in the literature \citep[e.g.][]{2004ApJ...609..482K,2009arXiv0903.4681M}, both distributions are equally consistent with the current dwarf distribution. This surprising result can be explained by a peculiar phase distribution of the orbits. In Figure \ref{fig:radialav}, the radial distribution of $z^{50}_0$ and $z^{50}_{10}$ averaged over the last 2 Gyrs are shown. On this time scale the orbits do not change significantly, but their phases are averaged out. Now we see a distinct difference between the two distributions. Therefore the apo-/pericenter distributions of both samples need to be compared with the dwarfs to differentiate between different dwarf formation scenarios. (Note that previous works using the same VL1 simulation that we use here did see a difference in the radial distribution of dwarfs selected at $z=0$ and those selected at $z=10$. This is because they did not use the same radial and mass cuts that we employ here. For real data, we may be lucky and see a biasing in the radial distribution, but to be certain to see it, we must use the apo/pericentre distributions.)

In Figure \ref{fig:dwarfs}, we show our derived orbits for nine Milky Way dwarf galaxies with observed proper motions. Analogous to the VL1 satellites (c.f. \S\ref{sec:measure}), we estimate our errors by building an ensemble of 1000 orbits for each dwarf drawn from its error distribution. The error bars show the values for the oblate L05 potential (blue) and for the TF model (red). Overlaid are the mean and standard deviation of the values from VL1 for the 50 most massive subhalos at $z=0$ and for the 50 most massive before $z=10$, respectively (black dashed lines and grey bands). The blue dashed lines denote the recovered mean of the $z^{50}_0$ and $z^{50}_{10}$ distributions, respectively (c.f. Figure \ref{fig:measureF}). Note that these lines indicate a tendency, as they only represent one possible realisation of the subhalo distribution. The galaxies are ordered by their observed star formation histories as described in \S\ref{sec:thedata}.

We do not model the LMC and SMC together as is often done in the literature \citep[e.g.][]{2007ApJ...668..949B,2008ApJ...684L..87B}, but investigate instead their orbits independently of each other. As shown in \S\ref{sec:3D}, our model systematics are too large to determine whether or not it is better to include LMC-SMC interactions in our models. Notice, that we only find bound orbits for the LMC because of the large Milky Way masses in our models. Recent results for LMC orbits in a lower MW mass potential by \citet{2006ApJ...652.1213K} and \citet{2007ApJ...668..949B} suggest that the LMC (and with it the SMC) are falling into the Milky Way for the first time. We find 6 satellites in our $z^{50}_0$ sample that fall into the main halo for the first time. Therefore we expect roughly one out of the nine dwarfs to be behave likewise. This makes a first infall scenario for the LMC not unlikely. 

Leaving aside the three most massive satellites in our sample -- Sagittarius, the SMC and the LMC -- the rest of the galaxies share a similar dynamical mass (see Table \ref{tab:dwarfdata1}). This makes them useful for probing how environment can promote or inhibit star formation. In Figure \ref{fig:dwarfs}, the dwarfs are ordered by their star formation histories (c.f. sec. \ref{sec:thedata}). We do not see evidence of enhanced star formation at pericenter, but rather a hint that it is suppressed. This agrees with \cite{2007Natur.445..738M}, who argue that low pericenters are a result of early accretion. Therefore dwarfs that are accreted earlier have suppressed star formation because of the higher UV background and lower pericenters. However, Gaia quality data will be required to test this convincingly.

Our recovered mean pericentre for all nine dwarfs is higher than the mean of both our $z^{50}_0$ and $z^{50}_{10}$ samples from the VL1 simulation. This could hint at satellite depletion by the galactic disc as was recently discussed in \citet{2009arXiv0907.3482D}. However, proper motion errors will bias our pericentre measurement to be systematically high (see further discussion below). Our derived apocentre distances $r_a$ are lower than the mean of the $z^{50}_0$ VL1 subhalos (the 50 most massive at redshift $z=0$). This discrepancy cannot be explained by a bias due to the proper motion errors as this effect has the wrong sign. Interestingly, however, the mean of our $z^{50}_{10}$ sample (the 50 most massive before redshift $z=10$) agrees well with the observed mean. This lends further support to the idea that the Milky Way's dwarfs formed early before reionisation \citep[e.g.][]{1992MNRAS.256P..43E,1999ApJ...523...54B,2000ApJ...539..517B,2002MNRAS.333..177B,2004ApJ...609..482K,2005ApJ...629..259R,2005MNRAS.364..367D,2006ApJ...645.1054G,2006MNRAS.368..563M,2009arXiv0903.4681M}.  The mean recovered $r_p/r_a$ ratio is higher in the dwarfs than in the simulation samples. Analogously to the pericenter distribution this can be explained by the bias from the proper motion errors. 

We investigate the influence of satellite depletion by a disc using the three subsets $z^{50}_{10}(r_d=10$\,kpc), $z^{50}_{10}(r_d=15$\,kpc) and $z^{50}_{10}(r_d=20$\,kpc) (50 most massive before $z=10$, without orbits that have pericenters $r_p <r_d$; \S\ref{sec:tests}). These explore difference effective disc sizes. Figure \ref{fig:dwarfs2} shows the same plots as Figure \ref{fig:dwarfs} for the the disc depleted samples. We find a trend to higher mean pericentres with larger effective disc radii, but also the mean apocenter increases. This is due to similar $r_p/r_a$ ratio in all three samples. Excluding Sagittarius, which is in the process of being disrupted by the disc, the apocenter distributions in $z^{50}_{10}(r_d=20$\,kpc) and $z^{50}_{10}(r_d=15$\,kpc) are too high in comparison to the current dwarf distribution. This suggests that the total destruction of satellites within $r_d > 15$\,kpc is too extreme. With the current data it is not possible to discriminate between the $z^{50}_{10}$ sample without disc depletion and $z^{50}_{10}(r_d=10$\,kpc) with a disc with radius $r_d=10$\,kpc.

\section{Conclusions}\label{sec:conclusions}
We have evaluated how well we can recover the orbits of Milky Way satellites in the light of measurement errors and model limitations. To do this, we compared orbits in a high resolution cosmological simulation of a Milky Way analogue with similar orbits integrated in a fixed background potential. We find that:

\begin{enumerate}
\item With current measurement errors, we can recover the last apocentre $r_a$ and pericentre $r_p$ to $\sim$40\%  ;
\item With Gaia quality proper motion data, we can recover $r_a$ and $r_p$ to $\sim$14\%, respectively, and the orbital period $t$ backwards over two orbits. In this regime, we become limited by model systematics rather than measurement error. In particular, how well we can approximate the shape of the Milky Way potential and how strongly orbits are affected by satellite-satellite interactions in infalling loose groups. 
\item recovering full {\it 3D} orbits -- the 3D pericentre ${\bf r_p}$ and apocentre ${\bf r_a}$, remains extremely challenging. This is due to a strong dependence on the potential shape that changes over time.
\end{enumerate}

We applied our orbit recovery technique to nine Milky Way dwarfs with observed proper motions to determine their last pericentre and apocentre distances $r_p$ and $r_a$. We found:

\begin{enumerate}
\item The mean recovered apocentres are lower than the mean of the most massive simulation subhalos at redshift $z=0$, but consistent with the mean of the most massive that form before $z=10$. This lends further support to the idea that dwarfs formed early before reionisation. 
\item With the current data a clear relation between star formation history and environment cannot be established. 
\end{enumerate}

\section*{Acknowledgments}

The authors would like to thank J. Guedes for providing her VL1 triaxiality data.

\bibliographystyle{mn2e}
\bibliography{OrbitInt_v1}

\end{document}